%% file: main.tex
\documentclass[12pt,journal]{IEEEtran}

\usepackage{moreverb,url}

\usepackage[colorlinks,bookmarksopen,bookmarksnumbered,citecolor=red,urlcolor=red]{hyperref}

\newcommand\BibTeX{{\rmfamily B\kern-.05em \textsc{i\kern-.025em b}\kern-.08em
T\kern-.1667em\lower.7ex\hbox{E}\kern-.125emX}}

\usepackage[numbers,sort&compress,square]{natbib}
\usepackage{ifpdf}
\ifpdf%                                % if we use pdflatex
  \pdfoutput=1\relax                   % create PDFs from pdfLaTeX
  \usepackage{graphicx}                % allow us to embed graphics files
  \DeclareGraphicsExtensions{.pdf,.png,.jpg,.jpeg} % for pdflatex we expect .pdf, .png, or .jpg files
\else%                                 % else we use pure latex
  \usepackage{graphicx}                % allow us to embed graphics files
  \DeclareGraphicsExtensions{.eps}     % for pure latex we expect eps files
\fi%
\graphicspath{{figures/}{pictures/}{images/}{./}} % where to search for the images
\usepackage{microtype}                 % use micro-typography (slightly more compact, better to read)
\PassOptionsToPackage{warn}{textcomp}  % to address font issues with \textrightarrow
\usepackage{textcomp}                  % use better special symbols
\usepackage{mathptmx}                  % use matching math font
\usepackage{times}                     % we use Times as the main font
\usepackage{tabu}                      % only used for the table example
\usepackage{booktabs}                  % only used for the table example
\usepackage{subfigure}
\usepackage{enumitem}
\usepackage{amsmath}
\usepackage{color}
\usepackage[final]{changes}
\definechangesauthor[color=blue]{VY}
\definechangesauthor[color=orange]{HW}
\definechangesauthor[color=purple]{HP}
\usepackage{xspace}

\newcommand{\move}[1]{}
\usepackage{amssymb}
\usepackage{rotating}

\title{Exploring the Limits of Complexity: A Survey of Empirical Studies on Graph Visualisation}

\author{Vahan Yoghourdjian, Daniel Archambault, Stephan Diehl, Tim Dwyer,\\Karsten Klein, Helen C. Purchase, and Hsiang-Yun Wu}

\begin{document}

\maketitle

\section*{Abstract}
For decades, researchers in information visualisation and graph drawing have focused on developing techniques for the layout and display of very large and complex networks. Experiments involving human participants have also explored the readability of different styles of layout and representations for such networks.  In both bodies of literature, networks are frequently referred to as being `large' or `complex', yet these terms are relative.  From a human-centred, experiment point-of-view, what constitutes `large' (for example) depends on several factors, such as data complexity, visual complexity, and the technology used. In this paper, we survey the literature on human-centred experiments to understand how, in practice, different features and characteristics of node-link diagrams affect visual complexity.
%the limits of scalability for node-link diagrams in practice in terms of raw numbers (e.g. nodes, links, and others). Based on our findings we give recommendations for future study design and reporting in network visualisation evaluation.

\section*{Keywords}
Graph Visualisation, Network Visualisation, Node-link Diagrams, Cognitive Scalability, Evaluations, Empirical Studies.

\input{sections/intro.tex}
\input{sections/relwork.tex}
\input{sections/methodology.tex}
 \input{sections/metrics.tex}
\input{sections/other-factors.tex}
 \input{sections/size-rationale.tex}
\input{sections/history.tex}

 \input{sections/conclusion.tex}

\section*{Acknowledgments}
This survey began as part of a working group output of the NII Shonan Seminar No. 2015-1 \textit{Big Graph Drawing: Metrics and Methods}, and we would like to thank this seminar series for the role it played in this survey. We would like to thank Tamara Munzner for her ideas and feedback at this seminar which helped focus the topic of this paper. The second author would like to thank EPSRC First Grant EP/N005724/1. The last author would like to thank the European Union’s Horizon 2020 research and innovation programme under the Marie Sklodowska-Curie grant agreement No 747985. This work was supported by the Australian Research Council Discovery Project grant DP140100077.

\bibliographystyle{abbrv} 
\bibliography{main,survey}

\end{document}

%% file: sections/intro.tex
%!TEX root = ../main.tex 
\section{Introduction}
There has been much work done on designing algorithms that can efficiently scale to create pictures of very large graphs. However, what remains a more open question, is whether pictures of very large and complex networks require a mental effort that exceeds the capabilities of an average human brain.

Eick~and~Karr~\cite{eick2002visualscalability} define the term \emph{visual scalability} as the capability of visualisation tools to effectively display large data sets. They also discuss factors affecting visual scalability, like human perception, monitor resolution, visual metaphors, interactivity, data structures and algorithms, as well as the computational infrastructure. A more recent discussion of these, and similar factors, are presented by Jankun-Kelly~\emph{et al.}~\cite{Kerren2014ScalabilityConsiderations}. They also distinguish perceptual and cognitive scalability: ``though elements may be perceivable, they may still exhaust cognitive resources.''

In cognitive psychology, Miller's `seven plus or minus two'~\cite{miller1956magical} is commonly accepted as a rule-of-thumb for the limitation on peoples' working memory.  Working memory is an example of one of the `cognitive ceilings' that might affect peoples' ability to reason about large networks.  Do working memory and other cognitive limitations have implications for the size and complexity of graphs that we should be trying to visualise?  

Huang~\emph{et al.}~\cite{IVJ:7} propose a framework for cognitive load in the context of graph visualisation. Based on results from cognitive psychology, they sketch a model that relates cognitive load during graph analysis tasks to mental effort and task performance in terms of response time and accuracy.  
They discuss a number of other factors affecting cognitive load: the domain (e.g. a highly technical and specific domain requires the user to relate their knowledge to the visual); the data itself (e.g. structure of the graph); the task (does it require deep understanding of the graph structure?); visual representation (does it follow best practice layout and design principles?); demographic (e.g.\ the experience of the users); and time pressure.  They report on one study of cognitive load that confirms some of these effects, but their model which suggests step changes in performance due to load, while compelling, is not fully validated.

In a more general evaluation of the effects of display types on visualisation cognition, Yost and North~\cite{Yost2006perceptualscalability} demonstrated that more pixels make it possible to show more data without considerable loss of performance. Going from 2 to 32 mega pixels led to a 20-fold increase in displayed data: the task completion times tripled, while accuracy only decreased from 95\% to 92\%. However, the data, visualisations and tasks considered (such as search and comparison) are relatively simple compared to networks and associated tasks involving understanding of connectivity.  It cannot be assumed that such results carry over, or that display size is the only---or even most---significant limitation.
The goal of this survey is to better understand these cognitive limitations.  To differentiate this human aspect of scalability of node-link diagrams from technological or technique specific limitations, we use the term \emph{cognitive scalability}.

This topic is important for our field, as such insights can guide the design of future techniques. For example, we are attempting to find tacit knowledge in past studies concerning the numbers of nodes and edges that are too difficult to work with in a single view.  If we can establish such numbers, then it might suggest that we need to direct efforts away from algorithms and rendering techniques that can scale to huge numbers of network elements.  Instead, for such large networks, we could focus on interactive ways to explore neighbourhoods (e.g.\ \cite{dwyer2008exploration,09vanHam}) or abstractions (e.g.\ \cite{06Abello,08ArchambaultGF,CHI:2}) instead of attempting to display the full set of nodes and links.

Thus, we survey a large number of papers reporting empirical studies of node-link diagrams, being exhaustive within the corpora of core visualisation proceedings and journals.  We aim to establish a consensus for definitions of adjectives like `large' or `dense' for node-link diagrams that are too complex to be easily comprehensible or useful for standard graph analysis tasks.  We also provide an overview of the types of networks and tasks, as well as experimental design of these experiments. 

In general, we find that the limits of scalability of the node-link network visualisation paradigm are rarely addressed directly.  Rather, there seem to be tacit assumptions (or possibly unreported pilot findings) about what size node-link diagrams are usable for different tasks, and experiments stay within these bounds while testing specific techniques.

Our key findings are that only a small range of graph sizes and structures have been used in experimental evaluations of graph visualisation techniques, mostly limited to small and sparse graphs. In particular, three quarters of studies use graphs with 100 nodes and 200 edges or less and, the remaining studies test interactive techniques, such that only a small portion of the graph is shown on the screen at a time. 
These findings are discussed further throughout the paper and listed in full in the conclusion.

The paper is structured as follows: Section~\ref{sec:relwork} discusses related surveys, primarily in graph visualisation; Section~\ref{sec:methodology} outlines our scope, methodology, and describes our categorisation framework; then we present the results of our survey in Sections \ref{sec:basic-metrics}, \ref{sec:other-factors} and \ref{sec:size-rationale}, followed by a discussion on trends in the network visualisation community in Section~\ref{sec:history}. 

%% file: sections/relwork.tex
%!TEX root = ../main.tex 
\section{Related Work}
\label{sec:relwork}

In the fields of graph drawing and visualisation, a number of surveys have considered scalability from different perspectives.  In particular, there has been much discussion of the scalability of algorithms and computer hardware to compute node-link diagram layout. Such papers tacitly acknowledge that very `large' and `dense' graphs are difficult to read and hence propose interaction techniques to navigate aggregated graphs. They may even report on studies of the readability of networks using different layout, interaction or rendering techniques. Yet, rarely do they explicitly address the question of what is the largest (most complex) diagram that people can usefully comprehend.  

Many past surveys characterise the techniques available for graph visualisation.  The surveys of Herman~\emph{et al.}~\cite {00Herman} and von Landesberger~\emph{et al.}~\cite {10STAR} are both of this type, focusing on techniques for graph visualisation and their strengths and weaknesses. Elmqvist and Fekete~\cite{09Elmqvist} characterise techniques in information visualisation that use hierarchical representations as a form of data abstraction.  Recent surveys have also focused on specific areas of network visualisation including multi-faceted graph visualisation~\cite{15Hadlak}, group structures in graphs~\cite{15Vehlow,vehlow2017visualizing}, matrix reordering techniques~\cite{surveyMatrixReorder}, edge bundling~\cite{surveyEdgeBundling}, and networks in social media~\cite{surveySocialMedia}.  

These surveys organise graph visualisation methods at the technique level, or specialise in a particular technique and present a survey of research in the area in-depth. These surveys do not focus on questions about how scalable these representations are from a human-centred perspective.  
%We contribute a survey from the perspective of experiments involving node-link diagrams in order to gauge the \emph{cognitive scalability} of these representations for both static and dynamic graphs. 

In the area of dynamic graphs, surveys have been conducted on dynamic networks~\cite {17Beck} and dynamic data in information visualisation in general~\cite {17Bach}. There have also been reviews focused on the human-centred effectiveness of animation, small multiples, and drawing stability (mental map preservation) by summarising experimental results and providing guidelines for visualisation designers. One such work by Archambault and Purchase~\cite{mapInMentalMap} summarises empirical results that relate to mental map preservation in dynamic graph drawing. In a later work~\cite{canAnimationSupport}, based on the results of new studies, they review the conditions where animation and small multiples are effective and present new results for diagrams of low drawing stability.  
%~\cite{mapInMentalMap,canAnimationSupport, more?}.  
These papers focus on dynamic network visualisation and do not consider network visualisation in general.  While providing a survey of human-centred effectiveness of visualisations to some degree, they do not consider cognitive scalability of the representations directly.  

In this paper, we review evaluations of node-link visualisations of static and dynamic graphs. What is unique to our survey is that it examines cognitive scalability of node-link visualisations of graphs through the lens of human-centred experiments, to gain bounds on the sizes of graphs that have been displayed to the human while still usefully supporting analysis tasks.  We seek to answer this question by surveying the literature of controlled experiments involving human participants to test node-link diagram representations of networks.  Our summary information about the networks, techniques and tasks considered in these studies also presents an up-to-date snapshot of the evolution and state-of-the-art of controlled network visualisation evaluation.

%% file: sections/methodology.tex
%!TEX root = ../main.tex 
\section{Methodology}
\label{sec:methodology}

In this section, we clarify the scope of this survey, including the venues that we examined, and describe our categorisation framework.  We have tried to be as systematic as possible in covering complete conference and journal venues with clearly defined constraints, as detailed below.

\subsection{Scope of Survey}

To the best of our ability, we have sought to include in this survey all papers with a human-centred experiment (formal user study) where at least one of the conditions is a node-link representation.  Both static and dynamic graph drawing studies were considered.
For much of our analysis we focus on individual studies, where papers could contain multiple studies. 
Throughout the survey, `paper' refers to the publication that presents the study, while `study' refers to an individual experiment. 

\begin{table}[ht]
\caption {Venues considered in this survey.
}\label{venues:tab}
\centering
{\scriptsize
    \begin{tabu} to \columnwidth {p{1cm}cccX}%
    \toprule
     \textbf{Venue} & \textbf{First Paper} & \textbf{\# of Studies} & \textbf{\# of Papers} & \textbf{References}\\
    \midrule
    \textit {ACM CHI} & 2006%1995 
    & 16 & 13 & \cite{CHI:1,CHI:2,CHI:3,CHI:4,CHI:5,CHI:6,CHI:7,CHI:8,CHI:9,CHI:10,CHI:11,CHI:12,CHI:13}\\
    %APVIS & ??? & 1 & \cite{APVIS:1} \\
    %CGF & ??? & 1 & \cite{CGF:1} \\
    \textit{Diagrams} & 2006 & 5 & 5 & \cite{DIAGRAMS:1,DIAGRAMS:2,DIAGRAMS:3,DIAGRAMS:4,DIAGRAMS:5}\\
    \textit {EuroVis} \& \textit{CGF} & 2009 & 16 & 13 & \cite{EUROVIS:1,EUROVIS:2,EUROVIS:3,EUROVIS:4,EUROVIS:5,EUROVIS:6,EUROVIS:7,EUROVIS:8,EUROVIS:9,EUROVIS:10,EUROVIS:11,CGF:1,CGF:2} \\
    \textit{GD} & 1995 & 25 & 23 & \cite{GD:1,GD:2,GD:3,GD:4,GD:5,GD:6,GD:7,GD:8,GD:9,GD:10,
                                     GD:11,GD:12,GD:13,GD:14,GD:15,GD:16,GD:17,GD:18,GD:19,GD:20,GD:21,GD:22,GD:23} \\
    \textit {IVJ} & 2002 & 19 & 14 & \cite{IVJ:1,IVJ:2,IVJ:3,IVJ:4,IVJ:5,IVJ:6,IVJ:7,IVJ:8,IVJ:9,IVJ:10,IVJ:11,IVJ:12,IVJ:13,IVJ:14} \\
    \textit{PacificVis} & 2008 & 7 & 6 & \cite{PACIFICVIS:1,PACIFICVIS:2,PACIFICVIS:3,PACIFICVIS:4,PACIFICVIS:5,PACIFICVIS:6}\\
    \textit{InfoVis} \& \textit {TVCG}  & 2003 & 61 & 47 & \cite{TVCG:1,TVCG:2,TVCG:3,TVCG:4,TVCG:5,TVCG:6,TVCG:7,TVCG:8,TVCG:9,TVCG:10,
                            TVCG:11,TVCG:12,TVCG:13,TVCG:14,TVCG:15,TVCG:16,TVCG:17,TVCG:18,TVCG:19,TVCG:20,
                            TVCG:21,TVCG:22,TVCG:23,TVCG:24,TVCG:25,TVCG:26,TVCG:27,TVCG:28,TVCG:29,TVCG:30,
                            TVCG:31,TVCG:32,TVCG:33, TVCG:34,TVCG:35, TVCG:36, TVCG:37, TVCG:38, TVCG:39, TVCG:40, TVCG:41, TVCG:42, TVCG:43, TVCG:44, TVCG:45, INFOVIS:1,INFOVIS:2,INFOVIS:3} \\
    %VINCI & ??? & 1 & \cite{VINCI:1} \\
    Other &   & 3 & 3 & \cite{VINCI:1, APVIS:1, VLHCC:1} \\
\midrule
\textbf{Total} & & \textbf {152} & \textbf {124} &\\    
\bottomrule
\end{tabu}
}
\end{table}

Our time range begins with the earliest formal human studies in network visualisation of which we are aware~\cite{GD:18} and ends on the $31^{st}$ of March 2018.  We consider the major conferences and journals listed in Table~\ref{venues:tab}, and examine all publications from this date, or the founding of, the conference/journal. This date is based on the above limit and accessibility of the venue.  Well-known articles outside these venues are also included and listed under `Other'.

For most venues, we were able to read all titles and further examine the abstracts of papers that were relevant to our survey.  
The only exception was \emph{ACM CHI}, where there were far too many papers: \emph{CHI} accepts up to 600 papers in total each year.  Instead, we found relevant papers at \emph{CHI} using the \emph{HCI Bibliography}.\footnote{http://hcibib.org/bs.cgi}  A query of this database, limited to the CHI conference and using search terms `{\tt (network | graph) visuali* study}', returned 25 results, of which close inspection revealed 12 to be relevant, as detailed in Table~\ref{venues:tab}\footnote{We also provide our curated bibliography as an online Tableau story\\
{\scriptsize  \tt \url{http://vahany.com/media/networkSize.html}}}.

\subsection{Categorisation Framework}
\label{sec:categorisation}

Information about each paper was collected and coded according to a number of criteria.
Section~\ref{sec:basic-metrics} reports our findings on the sizes of graphs used in experiments. We identified the number of nodes and edges used within each study and computed the density of the graphs.  If the study was on dynamic graphs, we counted the number of timeslices.  We noted if this information was explicitly stated or whether it needed to be derived or inferred (only nodes, only edges, or both).  Exact numbers were sometimes unavailable, but we estimated the sizes based on figures of the stimuli (e.g.\ \cite{VLHCC:1, VINCI:1}). If authors provided the algorithm and parameters used to generate the graphs, we computed the size based on this information (e.g.\ \cite{CHI:9, PACIFICVIS:2, IVJ:10, TVCG:4}).

Section~\ref{sec:other-factors} discusses other factors we found relating to scalability and, as such, is divided into three parts:  HCI factors, graph drawing factors, and study design. The first part presents our findings on factors relative to human computer interaction and scalability.  We coded and described the types of tasks~\cite{lee2006task} and interaction~\cite{yi2007toward} used in each study.
Application areas---if any---and the challenges they pose were also collected and are discussed in this section.
The second part presents graph drawing factors related to scalability.  We gathered information about the types of graphs, whether they were static or dynamic, and if they had attributes.  We recorded information about the graph structure and noted if the data was real or generated.  We also coded the layout algorithm used in the experiment.
The third part discusses factors with respect to study design, including the number and nature of participants and whether the studies were within or between subject. In addition, we discuss the results of the study, and present the studies that are interesting in this data set.

Section~\ref{sec:size-rationale} presents information about how the authors of experiments decide how large a graph should be presented to a participant.  In particular, it discusses pilot studies and justification for using graphs of a certain size.
In Section~\ref{sec:history}, we discuss the evolution of studies and their design over time in our community.  Information about how studies have evolved and their venues is presented here also.
The final sections of this paper include a discussion, recommendations for our community, and a conclusion.

%% file: sections/metrics.tex
%!TEX root = ../main.tex 
\section{Basic Measures of Complexity}
\label{sec:basic-metrics}

In this section, we consider measures of size that can be used to describe the graphs used in studies. 

There are numerous measures that could be used when considering complexity of graph visualisations. However, we have found that in reports on graph study design and methodology, typically, few are considered.  The number of nodes is the most commonly reported measure. The number of edges and/or density, on the other hand, is less frequently provided - although it is known to be a significant factor \cite{IVJ:9}. More edges, inevitably leads to more edge crossings, which is known to affect readability \cite{purchase1997aesthetic}.  When the data changes over time (i.e.\ dynamic graphs) additional measures, such as number of timeslices, become similarly important to gauge complexity.

\begin{table*}
\caption{Summary of graph sizes used in usability studies of graph visualisation. We used $\frac{|E|}{|V|(|V|-1)}$ to calculate density, and $\frac{|E|}{|V|}$ to calculate linear density.
\label{tab:stats}}
\centering
	\begin{tabu} to \columnwidth {ccccccc}%
	\toprule \textbf{Measure}
     & \textbf{Minimum} & \textbf{Maximum} & \textbf{Average} & \textbf{Median} & \textbf{Lower Quartile} & \textbf{Upper Quartile}\\
	\midrule
	\textbf {nodes} & 2 & 113 M. & 457,242 & 49.5 & 18.5 & 116.5 \\
       \textbf {edges} & 1 & 1.8 B. & 7.87 M. & 73 & 23 & 242.5 \\
        \textbf {density} & 0.014 & 115.5 & 10 & 5 & 1.7 & 10\\
        \textbf {linear density} & 0.38 & 102.8 & 3.5 & 1.5 & 1 & 2.4\\
				\textbf {timeslices} & 2 & 15 & 6 & 6 & 3 & 7 \\
\bottomrule
\end{tabu}

\end{table*}

Since our interest is primarily in findings regarding the scalability of network visualisation, we report on the distribution of graph sizes considered within studies.
Table~\ref{tab:stats} summarises the minimum, maximum, average, median, and upper/lower quartiles for these metrics. 
While there is a large range across all these metrics, in each case the median is closer to the minimum.  This skewed distribution implies that the majority of studies use small graphs.

\begin{figure*}
\centering
\includegraphics[width=\linewidth]{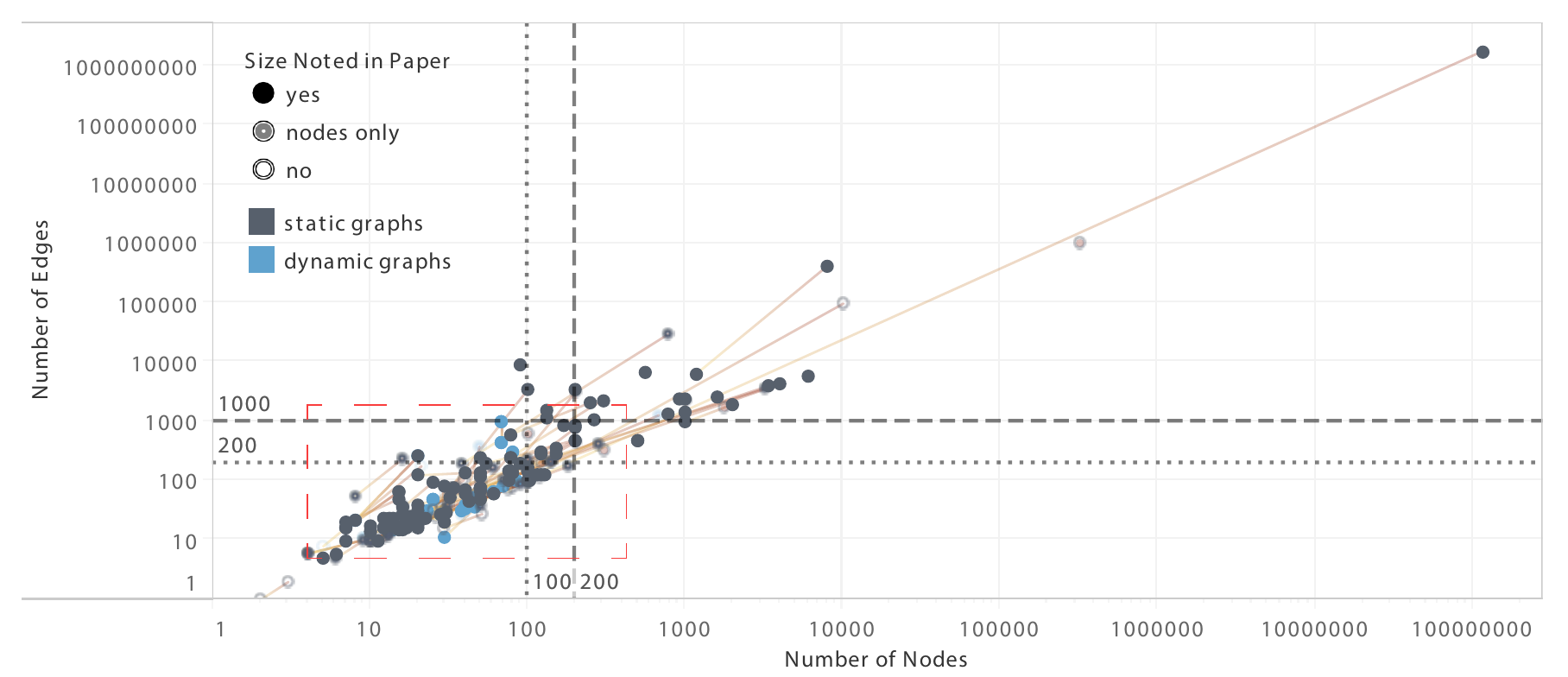}
\caption{\label{fig:graphsizes} The size of graphs used in user studies. The x-axis shows the number of nodes, while the y-axis shows the number of edges. Both axes have a log scale. The grey circles represent static graphs, while the blue show dynamic graphs.}
\end{figure*}

\begin{figure*}
\centering
\includegraphics[width=\linewidth]{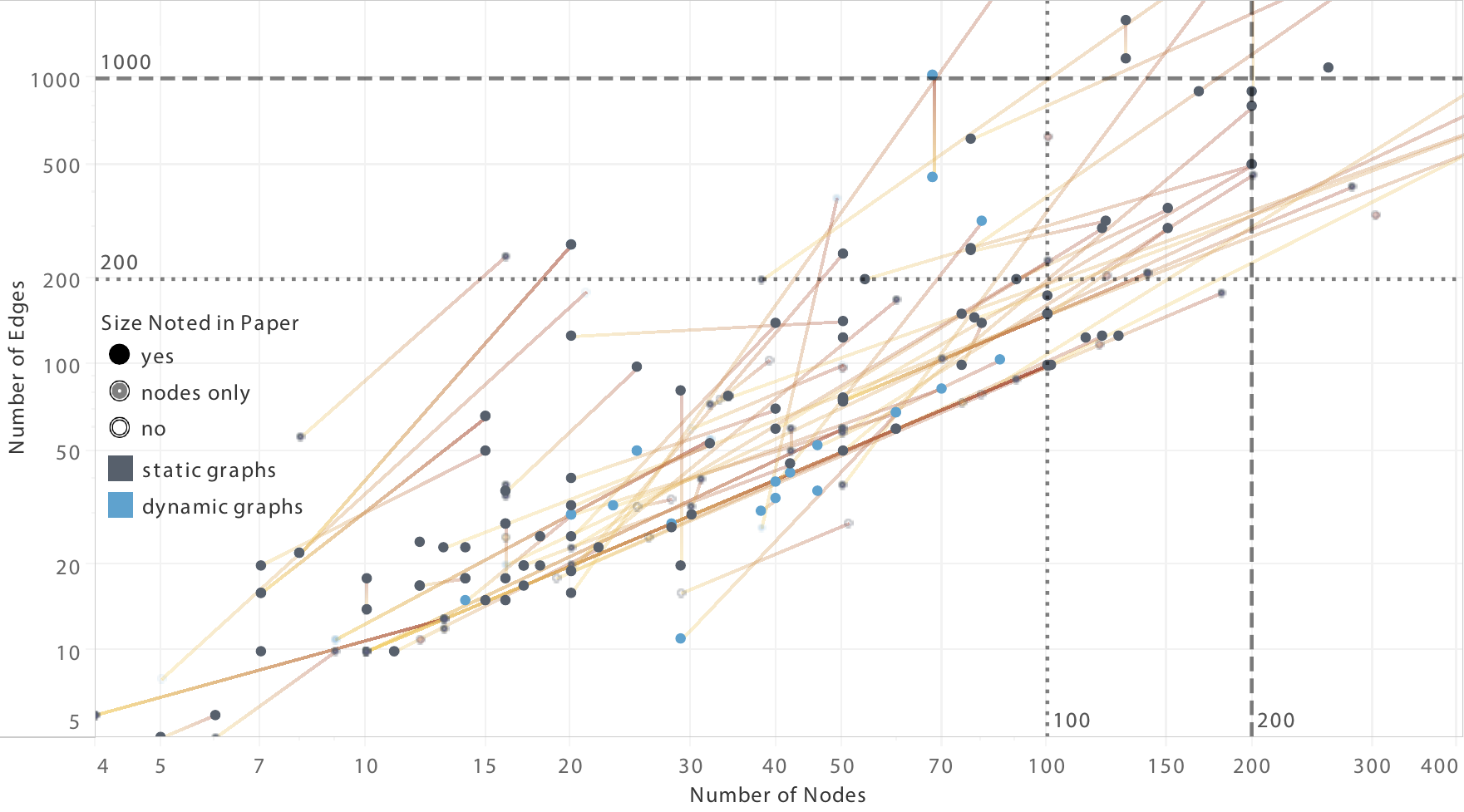}
\caption{\label{fig:graphsizes-zoom} The section, marked by red dotted lines in Fig.~\ref{fig:graphsizes}, magnified for better readability.}
\end{figure*}

In order to have a better understanding of the range of sizes of graphs used across and within user studies, we plotted for each study the number of nodes against the number of edges for both the smallest graph used in each study and the largest (Figs.~\ref{fig:graphsizes},~\ref{fig:graphsizes-zoom}). Each minimum and maximum pair is connected by a link showing the range of graph sizes evaluated in that study. 
The circles are: hollow \includegraphics[height=2.5mm,trim=0mm 0mm 0mm 0mm]{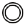} if the number of nodes and edges was not mentioned by the authors; double enclosed \includegraphics[height=2.5mm,trim=0mm 0mm 0mm 0mm]{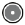} if only the number of nodes was mentioned; full \includegraphics[height=2.5mm,trim=0mm 0mm 0mm 0mm]{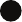} if both number of nodes and edges were mentioned, or both metrics were otherwise apparent (e.g.\ if number of nodes $n$ was given and the graph type was a tree, we assumed $n-1$ edges were present). In cases where the information was not mentioned we estimated the graph sizes from the figures (e.g.\ \cite{CHI:7,CHI:11,EUROVIS:4, EUROVIS:7, GD:3, TVCG:9, TVCG:12, TVCG:17, TVCG:20, TVCG:30, TVCG:32, CGF:1}).  
We also used the hue of the circles to differentiate between static \includegraphics[height=2.5mm,trim=0mm 0mm 0mm 0mm]{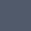} and dynamic \includegraphics[height=2.5mm,trim=0mm 0mm 0mm 0mm]{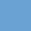} graphs.

Where precise sizes were not provided, we did our best to infer approximate sizes. For example, in some cases we were able to estimate the graph sizes from the figures (\cite{CHI:7,CHI:11,EUROVIS:4, EUROVIS:7, GD:3, TVCG:9, TVCG:12, TVCG:17, TVCG:20, TVCG:30, TVCG:32, CGF:1}).

The following subsections discuss each of the four measures of Table~\ref{tab:stats} in more detail. 

%-----------------------
%-----------------------
\subsection {Number of Nodes}
\label{sec:metrics-nodes}

\begin{figure*}
\centering
\includegraphics[width=\linewidth]{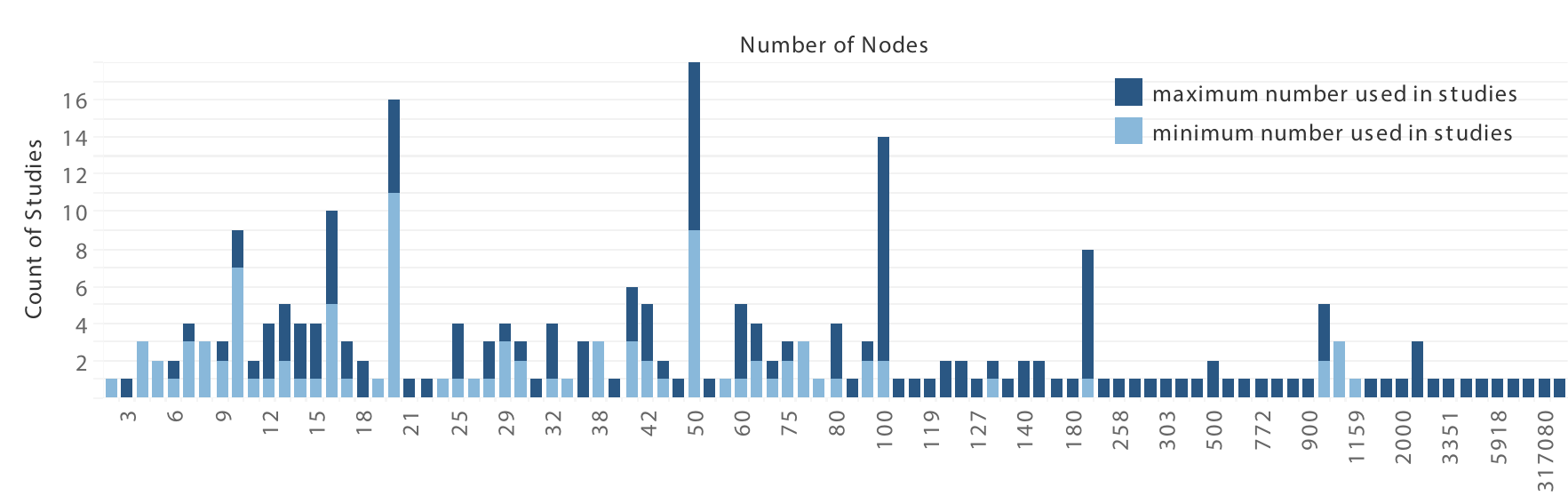}
\caption{
Histogram of the minimum (light blue) and maximum (dark blue) number of nodes of graphs used in studies.
\label{fig:graphnodes}}
\end{figure*}

The number of nodes in a graph is an important, but incomplete indicator, of complexity. 
It is clear that, for most tasks, difficulty is affected by introducing more nodes to a connected graph.  We would assume that most experimenters pilot, or at least consider, graphs with different numbers of nodes to avoid tasks that are too trivial or impossible.  Yet out of the 152 studies covered in this survey, 28 studies do not mention node count at all.  
%As mentioned above, for some of these studies, we were able to estimate the number of nodes from the figures provided in the paper.

Among the studies that are included in our survey, 15 studies use graphs with more than 1,000 nodes \cite{GD:10, TVCG:10, TVCG:8, CHI:12, CHI:2, EUROVIS:9, GD:6, IVJ:10, TVCG:29, TVCG:34, IVJ:11, TVCG:40, TVCG:41} and another nine that use graphs with more than 500 nodes \cite{CHI:4, CHI:6, EUROVIS:1, TVCG:13, TVCG:14, TVCG:2, TVCG:41, TVCG:38}.

Among these studies, the majority aim to evaluate tools that use interactive exploration to extract parts (e.g.\ neighbourhoods) of the graph (20 / 24 studies, 83\%). Some of these studies evaluate aggregation techniques, thus they would require graphs with a large number of nodes, in order to highlight the benefits of compressing several nodes into fewer representations.

It is problematic to infer cognitive scalability of graph visualisation in the presence of interactivity because most of these studies do not ask the participants to perform the tasks on the whole graph. Rather, only a part of the graph is visible.
When the authors do not report the precise number of nodes actually visible to the user, there is little that can be inferred about cognitive scalability.
Eight of these 24 studies also use smaller networks with fewer than 100 nodes in their evaluations. This is shown by the long lines (Fig.~\ref{fig:graphsizes},~\ref{fig:graphsizes-zoom}) that connect different graph sizes used by the same studies.

In order to understand the selection of number of nodes, we plotted a histogram of the number of nodes. Fig.~\ref{fig:graphnodes} shows a number of spikes around specific numbers of nodes: 20, 50, and 100.
The biggest spikes are at graphs with 50 nodes, which were used by 18 studies, followed by 20 and 100, used in 16 and 14 studies respectively.
We believe that spikes at these round numbers suggest experimenters choose the number of nodes arbitrarily. These round numbers were not identified by empirical research and a formal study on ceiling and floor effects for graph cognitive scalability might lead to a better selection of number of nodes.

Nine additional studies use graphs with more than 200 nodes. Similar to the above, some of these studies only show subparts of the network \cite{DIAGRAMS:5,EUROVIS:11,TVCG:25,GD:22}, while some evaluate tools that scale well with large networks \cite{DIAGRAMS:5,EUROVIS:11,TVCG:25,PACIFICVIS:2,GD:9,TVCG:9,PACIFICVIS:5,GD:19}. Four out of these nine studies have also used networks with less than 100 nodes.

To conclude, only 56 out of 152 studies use graphs with more than 100 nodes. Most of the studies that use a large number of nodes, use abstractions or aggregation to show only parts of the graph at a time, but do not report on the number of nodes seen at a given abstraction. In general, it is not clear why most researchers choose to use graphs with 100 nodes or less (121 / 152 studies, 80\%). Some authors mention pilot studies where they discovered ceiling or floor effects, which could be seen as indications of cognitive scalability. We discuss these in Section~\ref{sec:pilot}; however to our knowledge, controlled studies were not conducted to verify or explain these effects.

%-----------------------
%-----------------------
\subsection {Number of Edges}
\label{sec:metrics-edges}

\begin{figure*}
\centering
\includegraphics[width=\linewidth]{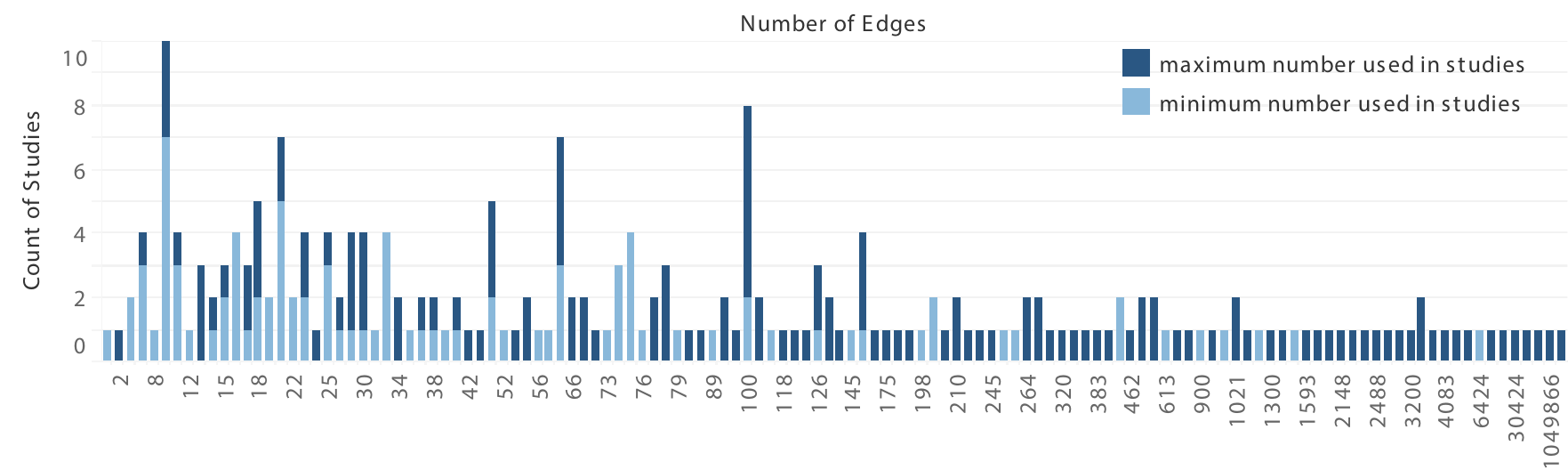}
\caption{Histogram of the minimum (light blue) and maximum (dark blue) number of edges of graphs used in studies.}
\label{fig:graphedges}
\end{figure*}

The number of edges in a graph is also an important indicator of complexity, especially in node-link diagrams, where the edges are drawn as lines. A large number of edges necessitates an increased number of crossings and overlap (collinear and therefore ambiguous lines). An excessive number of links often lead to what are known as `hairball' visualisations.
Among the 152 studies covered in this survey, 124 explicitly mention the number of nodes, while only 87 specify the number of edges.

There are only 28 studies that use graphs with 1,000 edges or more. Most of these allow the participants to look at parts of the network instead of performing the task on the whole network \cite{CHI:12,CHI:2,EUROVIS:9,IVJ:10,IVJ:11,IVJ:12,TVCG:21,EUROVIS:1,CHI:4,TVCG:25,GD:6,GD:9,GD:22,TVCG:29,TVCG:14,TVCG:31,TVCG:45,TVCG:10,TVCG:38,TVCG:24}. Similar to studies that use a large number of nodes to highlight the benefits of aggregation or interaction methods, some studies use a large number of edges to show the benefits of edge compression, bundling, or highlighting techniques. Many evaluate tools or techniques which, by design, scale well to handle graphs with a large number of edges. For example, a study by Giacomo~\emph{et al.}~\cite{GD:9} evaluates a technique that highlights edges in order to enhance the readability of graphs that have many edge crossings. Another category of studies that use a large number of edges, evaluate visualisations (e.g.\ adjacency matrices) that scale well with a large number of edges in comparison to node-link diagrams \cite{IVJ:9,TVCG:6,TVCG:31,TVCG:39,TVCG:40,GD:22}.

Again, just as we saw spikes in frequency at round numbers of nodes used for study graphs (Fig.~\ref{fig:graphnodes}), in Fig.~\ref{fig:graphedges} we see spikes particularly at 10, 20, 60, and 100 edges.
However, the spikes are less pronounced: only ten, seven, seven, and eight studies, respectively.  
The smaller spikes for number of edges compared to number of nodes, tends to suggest that experimenters choose a specific number of nodes first, then adjust the density, presumably to control the difficulty level for their specific tasks. 
We found that 97 studies use graphs with less than 100 edges, while 75 studies use graphs with 100 edges or more.

In summary, we were surprised that about half of the studies do not report the number of edges. We would argue that without this information, graph evaluations are difficult to reproduce.
The majority of studies use graphs with less than 1,000 edges (126 / 152 studies, 83\%). The 28 studies that use graphs with 1,000 edges or more, either evaluate tools that require dense community structures, aim to show that node-link diagrams fail to perform well on graphs with a large number of edges, or highlight the benefits of aggregation and interaction techniques. %which help visualisations scale to graphs with large number of edges.

%-----------------------
%-----------------------
\subsection{Density}
\label{sec:metrics-density}

The previous section mentions studies that use graphs with a large number of edges. Nonetheless, most of these graphs have very low densities (relatively few edges to the number of nodes).

In this section we focus on studies that explicitly consider the effect of density on readability.
We used $\frac{|E|}{|V|(|V|-1)}$ to calculate density. This represents the ratio of the number of existing edges $E$ to the number of all possible edges for the number of nodes $V$. For simplicity, we multiply this ratio by $100$ to achieve percentages.

There are also other ways to derive density. The so-called \emph{linear density} $\frac{|E|}{|V|}$ is often used to compare the number of edges $E$ to the number of nodes $V$.  In this measure, tree-like graphs will have density close to 1. Ghoniem~\emph{et al.}~\cite{INFOVIS:1} suggest using \emph{square-root density} $\sqrt{\frac{|E|}{|V|^2}}$, since its range is bounded to the interval $[0,\frac{1}{\sqrt{2}})$. Any of these definitions can be used; the choice depends on utility \cite{melancon2006just}.

\begin{sidewaysfigure*}
\centering
\includegraphics[width=\textwidth]{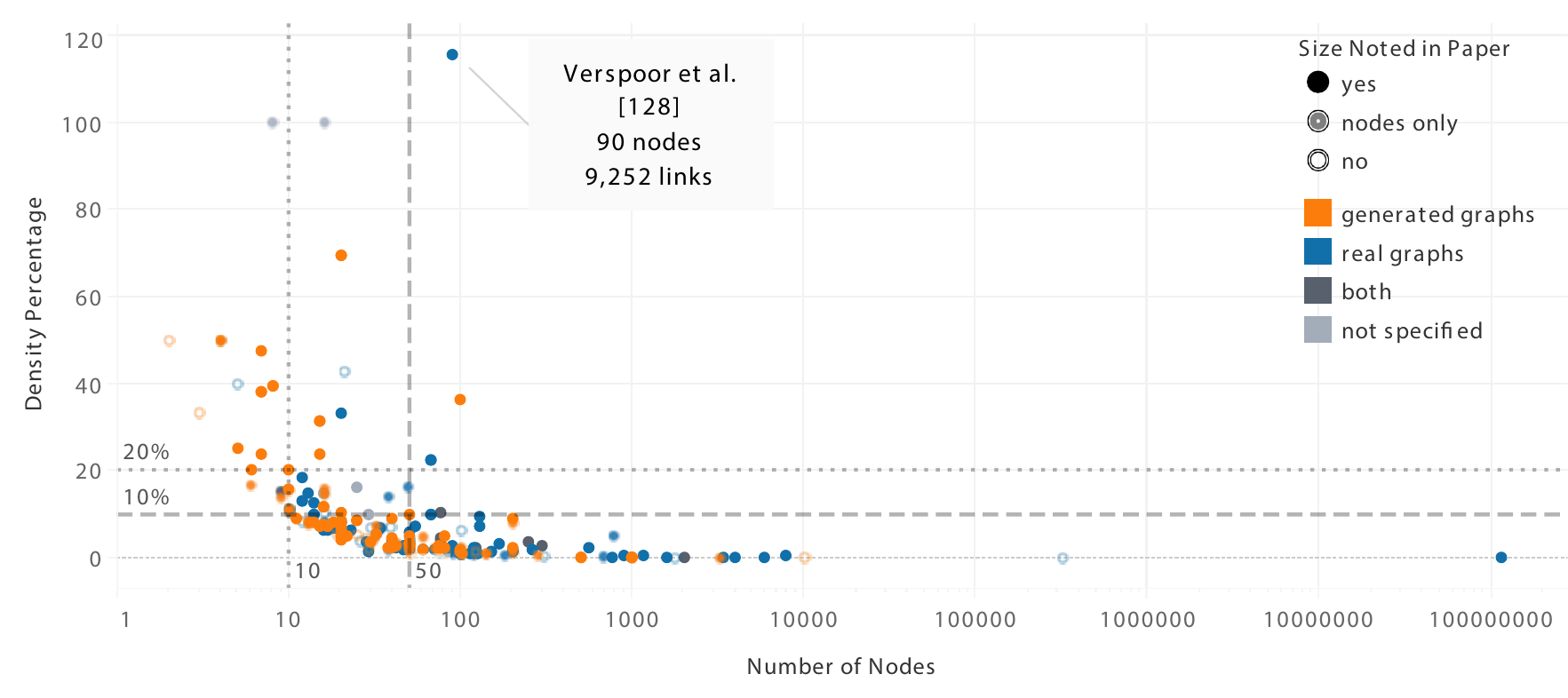}
\caption{\label{fig:graphdensity} The density of graphs used in the studies. Out of 152 studies, 129 studies use sparse graphs with densities of less than 20\%, while 119 studies use graphs with less than 10\% density.}
\end{sidewaysfigure*}

Fig.~\ref{fig:graphdensity} shows the density of real and generated graphs used in studies. Most studies %covered by this survey 
use graphs with densities of $50\%$ or less.
There are only three studies that use graphs with densities higher than $50\%$ \cite{IVJ:12,TVCG:36,TVCG:22}. 
All three studies evaluate methods that are tailored to scale well for dense networks.

In fact, all studies that use graphs with ten nodes or more and a density of more than 20\% evaluate matrix-like visualisations that scale well for dense graphs \cite{CHI:7,IVJ:6,IVJ:9,TVCG:39} or evaluate edge-compression and edge-bundling tools \cite{TVCG:23,TVCG:37}.

In summary, in the surveyed studies, dense graphs are mostly small and have less than ten nodes. The only studies that evaluate visualisations using dense graphs ($>20\%$ density) with more than a trivial number of nodes, tend to be those testing matrix diagrams or edge compression techniques. With the exception of these types of studies, all studies that use graphs with more than 50 nodes, choose to use sparse graphs with a density of less than 10\%. 119 out of 152 studies (78\%) use graphs with a density of less than 10\%, while 129 out of 152 studies (85\%) use graphs with less than 20\% density.
%-----------------------
%-----------------------
\subsection{Number of Timeslices}
\label{sec:metrics-timeslots}

In the case of dynamic graphs or static graphs with dynamic attributes, and in addition to the number of nodes, and the number of edges or density, the number of timeslices is an important measure of cognitive scalability. 

Among the 152 studies (described in 124 papers) covered in this survey, 22 use dynamic graphs.
Compared to the reporting of size metrics discussed above, it seems dynamic graph evaluation papers are reasonably consistent about reporting the number of timeslices.

\begin{figure}
\centering
\includegraphics[width=\columnwidth]{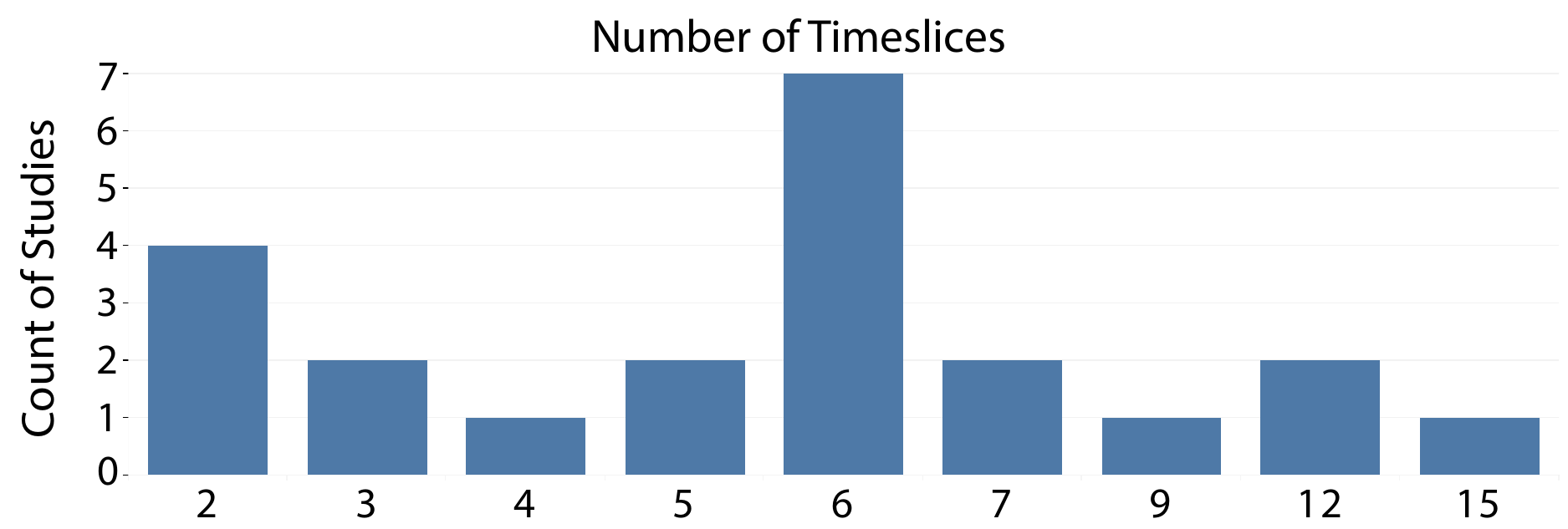}
\caption{\label{fig:timeslices-histogram} Histogram \replaced[id=VY]{of the number of timeslices of dynamic graphs used in the surveyed studies.}{of timeslices of dynamic graphs used.} }
\end{figure}

Fig.~\ref{fig:timeslices-histogram} shows a histogram of the number of timeslices. For previous metrics, we used the minimum and maximum values to derive the charts, but since there are only a few studies that use dynamic graphs, we included all the values. There is no visible difference between odd and even numbers. There is, however, a clearly visible spike at six timeslices. Seven studies, out of the total 22, use dynamic graphs with six timeslices.  

We suggest that six timeslices is typically chosen as it is small enough for a small multiples representation of a dynamic graph to fit on a screen with each timeslice being at a reasonable scale. 
Farrugia and Quigley~\cite{IVJ:5} conduct two studies in order to compare animated displays to static ones. They used two graphs with six timeslices each. For the static view, they placed the six timeslices next to each other on a 2$\times$3 grid.
Similarly, Archambault and Purchase~\cite{IVJ:8} conduct a study that evaluates the effects of three factors-- static versus animated presentation of dynamic attribute values, force-directed versus hierarchical layout of constant graph structure, and `with history' versus `without history' persistence, for displaying graphs with dynamic attributes. They use six timeslices, with each covering one-sixth of the screen.

Shi~\emph{et al.}~\cite{TVCG:24} conduct the only study that tests a variety of time slice counts. They demonstrate the scalability of various aggregation techniques against the more typical small multiples display of timeslices.
They use four dynamic graphs with different numbers of timeslices. They use graphs with 15, 12, 4, and 5 timeslices and 674, 109, 298, and 16 nodes respectively. 
For their small-multiples condition, they use either a 6 or 24-cell grid to make the full set of timeslices for each graph visible.  
Unlike the studies above, where the number of timeslices were chosen to fill the small multiplies grid, in this study the odd numbers of timeslices would have left part of the screen unused.

Others choose the number of timeslices according to the norms of specific application areas.
North~\emph{et al.}~\cite{IVJ:1} use a directed graph with 46 nodes, 36 edges, and 12 timeslices in their evaluation of three existing visualisation methods for dynamic graphs. They claim that this is a typical size of pathways used by Biologists.

In addition to the number of timeslices, the number of graph elements that change from one timeslice to another is important. Authors report these values in different ways. For example, some state the maximum number of changed elements \cite{GD:13}, while others report on the average \cite{DIAGRAMS:2, GD:11}.

In summary, the number of timeslices for dynamic graphs is often small. Most studies use less than ten timeslices (except for \cite{IVJ:1, TVCG:24}). Moreover, they often consider only a fixed number of timeslices (except Shi~\emph{et al.}~\cite{TVCG:24} who use four variations). Some studies pick the number of timeslices to best fit their visualisations on screen. Others choose what is common in the respective application area. 

%% file: sections/other-factors.tex
%!TEX root = ../main.tex 
\section{Other Factors and Scalability}
\label{sec:other-factors}

In addition to the basic measures of the previous section, other factors and their interplay influence the cognitive scalability of graph visualisations.
In the following sections, we discuss HCI, graph drawing, and study design factors.

\subsection {HCI Factors}

\begin{figure*}
\centering {
\tabcolsep=1pt
\begin{tabular}{cc}
\includegraphics[width=0.45\linewidth]{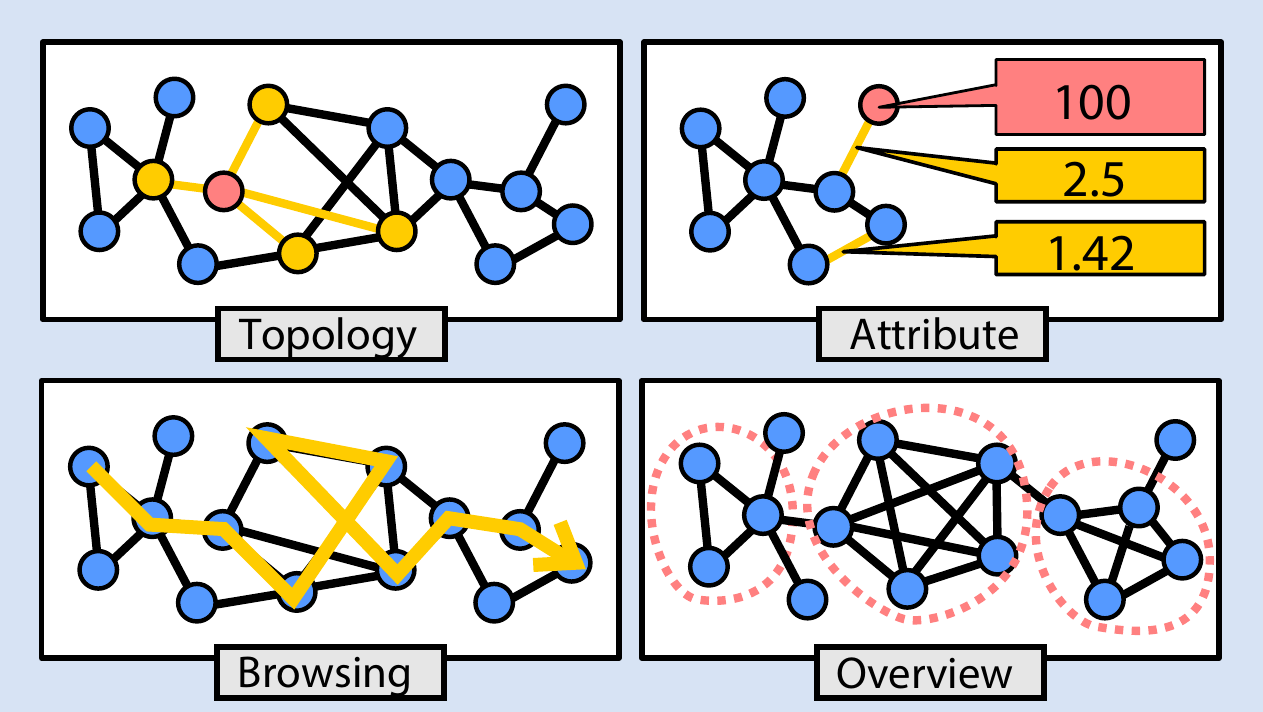} \\
\includegraphics[width=\linewidth]{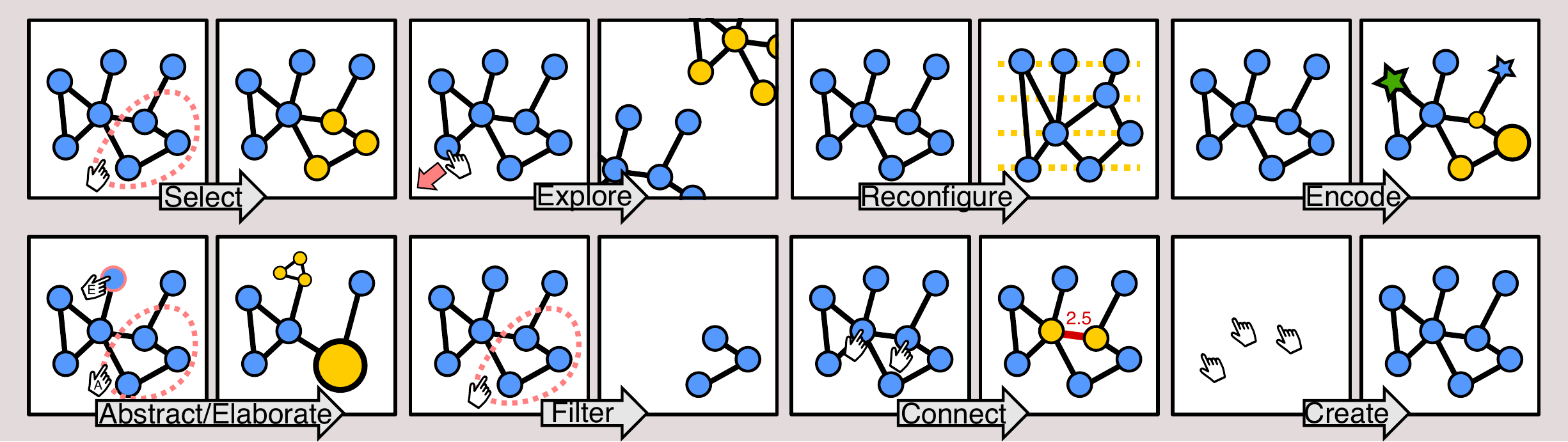} \\
(a) & (b) 
%\color{blue}(a) & \color{red}(b) 
\end{tabular}
}
\caption{\label{fig:tax-tasks-interactions} 
Task and interaction taxonomies. \color{black}Task taxonomy (a) proposed by Lee et al.~\cite{lee2006task}, which mainly includes \emph{Topology}, \emph{Attribute}, \emph{Browsing}, and \emph{Overview} tasks. \color{black}Interaction types (b) introduced by \emph{Yi et al.}~\cite{yi2007toward}, which include \emph{Select}, \emph{Explore}, \emph{Reconfigure}, \emph{Encode}, \emph{Abstract/Elaborate}, \emph{Filter}, and \emph{Connect}. However, \emph{Create} is newly added in our taxonomy.}
\end{figure*}

Graph visualisations are often tailored to efficiently serve domain-specific tasks. Modern visualisation tools are equipped with heavy interaction. Such Human-Computer Interaction (HCI) factors play a key role in allowing graph visualisations to scale to larger data sets. They were rarely investigated in earlier visualisations, but have been widely discussed in recent times.

In this section, we discuss the interplay between different types of tasks, interaction techniques and application areas on one hand, and the basic measures of graph size on the other.

\subsubsection {Tasks}
\label{sec:otherFactors:tasks}

%\begin {itemize}[leftmargin=*, itemindent=1em]
%\item \textbf{Tasks:}
To analyse and understand the trend, as well as the impacts of scalability in relation to tasks, we classified the various tasks used in studies into four main categories based on the taxonomy of Lee~\emph{et al.}~\cite{lee2006task}. Fig.~\ref{fig:tax-tasks-interactions}(a) shows a summary of tasks investigated in our survey: %\emph{Topology-based}, \emph{Attribute-based}, \emph{Browsing}, and \emph{Overview} tasks. 
\emph{Topology-based} tasks allow participants to detect node adjacency, accessibility, common connection, and connectivity. 
%This includes identifying paths~\cite{GD:9}, the length of the shortest path~\cite{GD:4, GD:20, PV:6, TVCG:12}, determining the degree of a node (e.g. \cite{IVJ:9}, find the most connected node?, \cite{TVCG:25, TVCG:21, IVJ:6, GD:15}, 
\emph{Attribute-based} tasks support identification of nodes and links by the data attributes associated with them.
%?which of the following nodes are adjacent to both given nodes?, \cite{CHI:3, CHI:4, IVJ:6, IVJ:8}. For dynamic graphs, the typical task is one of asking participants to identify changes in the graph structure (e.g. \cite{GD:8} ?When is the first appearance of node X??, \cite{GD:11, GD:6, EV:6}. 
%In some cases, the tasks are more than simple graph-reading tasks, and involve coordinated use of the whole graph.  \cite{TVCG:3, TVCG:17, TVCG:12} asked participants to create their own layout of a given graph, the aim being to see which layout principles users impose of graph drawing. \cite{TCVG:2} required that participants drew the graph from scratch (ensuring no initial layout bias), while \cite{TVCG:15} asked participants to sketch a graph drawing from memory. The additional complexity of these layout/drawing tasks mean that the maximum sized graph in these four studies is 50/77 \cite{TVCG:3}.
\emph{Browsing} tasks include following or revisiting a path. High-level or deliberately more abstract tasks were classified as \emph{Overview} tasks. Note that we considered tasks that asked the participants to find the shortest path to be in the \emph{Topology-based} category, even if they required some path following. We categorised path following tasks as \emph{Browsing}, only if participants were explicitly asked to follow a certain path.

\begin{table}
\caption {The type of tasks used within the studies.}\label{tab:task}
\centering
{\scriptsize
	\begin{tabu} to \columnwidth {cXc}%
	\toprule
     \textbf{Task} & \textbf{References} & \textbf{\# of Studies} \\
	\midrule
Attribute & \cite{CHI:1,CHI:2,CHI:6,CHI:7,CHI:8,CHI:10,CHI:11,CGF:1,DIAGRAMS:4,EUROVIS:2,EUROVIS:5,EUROVIS:6,EUROVIS:8,EUROVIS:9,GD:3,GD:5,GD:6,GD:8,GD:12,IVJ:1,IVJ:3,IVJ:6,IVJ:8,IVJ:9,INFOVIS:2,TVCG:8,TVCG:11,TVCG:13,TVCG:16,TVCG:20,TVCG:21,TVCG:22,TVCG:26,TVCG:27,TVCG:28,TVCG:30,TVCG:32,TVCG:33,TVCG:38,TVCG:43,TVCG:45,IVJ:13,IVJ:14,EUROVIS:10,EUROVIS:11} & 53 \\
Browsing & \cite{CHI:3,CHI:4,CHI:7,CHI:8,EUROVIS:2,EUROVIS:9,GD:13,GD:14,IVJ:10,PACIFICVIS:1,PACIFICVIS:3,TVCG:3,TVCG:5,TVCG:6,INFOVIS:3,TVCG:18,TVCG:21,TVCG:25,TVCG:38,TVCG:43,TVCG:44,IVJ:14,GD:21,CGF:2} & 26 \\
Overview & \cite{CHI:2,CHI:3,CHI:5,CHI:7,EUROVIS:5,EUROVIS:7,EUROVIS:9,GD:5,GD:6,GD:8,GD:10,GD:11,GD:12,GD:17,GD:19,IVJ:1,IVJ:9,IVJ:10,PACIFICVIS:4,TVCG:1,TVCG:5,TVCG:8,TVCG:11,TVCG:12,TVCG:13,TVCG:14,TVCG:20,TVCG:22,TVCG:24,TVCG:26,TVCG:29,TVCG:31,TVCG:34,TVCG:38,TVCG:39,TVCG:40,TVCG:41,IVJ:11,IVJ:12,IVJ:13,GD:22,GD:23} & 52 \\
Topology & \cite{CHI:2,CHI:3,CHI:4,CHI:5,CHI:6,CHI:7,CHI:8,CHI:9,CHI:10,CHI:12,APVIS:1,CGF:1,DIAGRAMS:1,DIAGRAMS:2,DIAGRAMS:3,DIAGRAMS:4,DIAGRAMS:5,EUROVIS:1,EUROVIS:2,EUROVIS:3,EUROVIS:4,EUROVIS:5,EUROVIS:7,EUROVIS:8,EUROVIS:9,GD:1,GD:2,GD:3,GD:4,GD:5,GD:7,GD:8,GD:9,GD:11,GD:12,GD:15,GD:16,GD:18,GD:20,IVJ:1,IVJ:2,IVJ:3,IVJ:4,IVJ:5,IVJ:6,IVJ:7,IVJ:9,IVJ:10,PACIFICVIS:1,PACIFICVIS:2,PACIFICVIS:3,PACIFICVIS:4,PACIFICVIS:5,PACIFICVIS:6,TVCG:1,TVCG:2,TVCG:3,TVCG:4,TVCG:5,TVCG:7,INFOVIS:2,TVCG:9,TVCG:10,TVCG:12,TVCG:13,TVCG:15,TVCG:16,TVCG:17,TVCG:19,TVCG:20,TVCG:21,TVCG:22,TVCG:23,TVCG:24,TVCG:25,TVCG:28,TVCG:31,TVCG:32,TVCG:33,VINCI:1,VLHCC:1,CHI:13,TVCG:35,TVCG:36,TVCG:37,TVCG:38,TVCG:39,TVCG:42,TVCG:43,TVCG:44,TVCG:45,IVJ:11,IVJ:12,IVJ:14,GD:21,GD:22,GD:23,CGF:2,EUROVIS:10,EUROVIS:11} & 114 \\

 \bottomrule
\end{tabu}
}
\end{table}

Table~\ref{tab:task} shows how the different tasks were used in the 152 studies (noting that several studies used more than one task). \emph{Topology-based} tasks were most common (47\%), followed by \emph{Attribute-based} (22\%) and \emph{Overview} (21\%).
Three quarters of the studies (114 studies) used tasks that are categorised as \emph{Topology-based}.

This might be an indication that researchers perceive tasks concerned with understanding the structure of the graph as good evaluation measures for graph visualisation. However, the number of studies that use \emph{Topology-based} decreases as graph size increases, while \emph{Overview} tasks become more popular. They become equally popular (37\%) in studies that use graphs with 300 nodes or more. For graphs with 1,000 nodes or more, \emph{Overview} tasks are used excessively (43\%), while \emph{Topology-based} tasks become less common (30\%). We believe that this is expected, since \emph{Overview} tasks do not require a detailed understanding of the graph, but deal with general properties and estimates.

\begin{sidewaysfigure*}
\centering
\includegraphics[width=\linewidth]{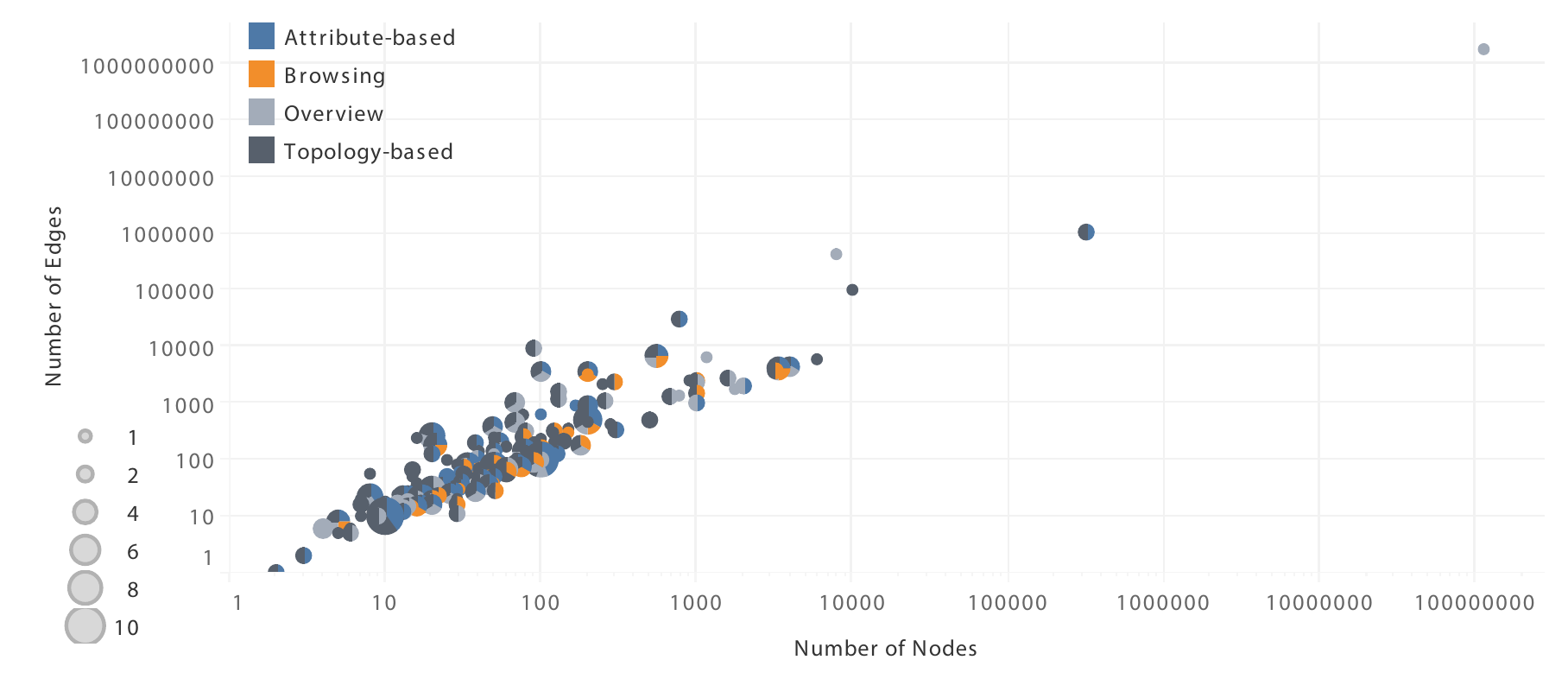}
\caption{\label{fig:tasks-percentage} Types of different categories of tasks used in studies with relation to the number of nodes (on the x-axis) and the number of edges (on the y-axis). The areas of the circles reflect the count of studies that use similar values. Both axes have log scales.}
\end{sidewaysfigure*}

\begin{figure*}
\centering
\includegraphics[width=\linewidth]{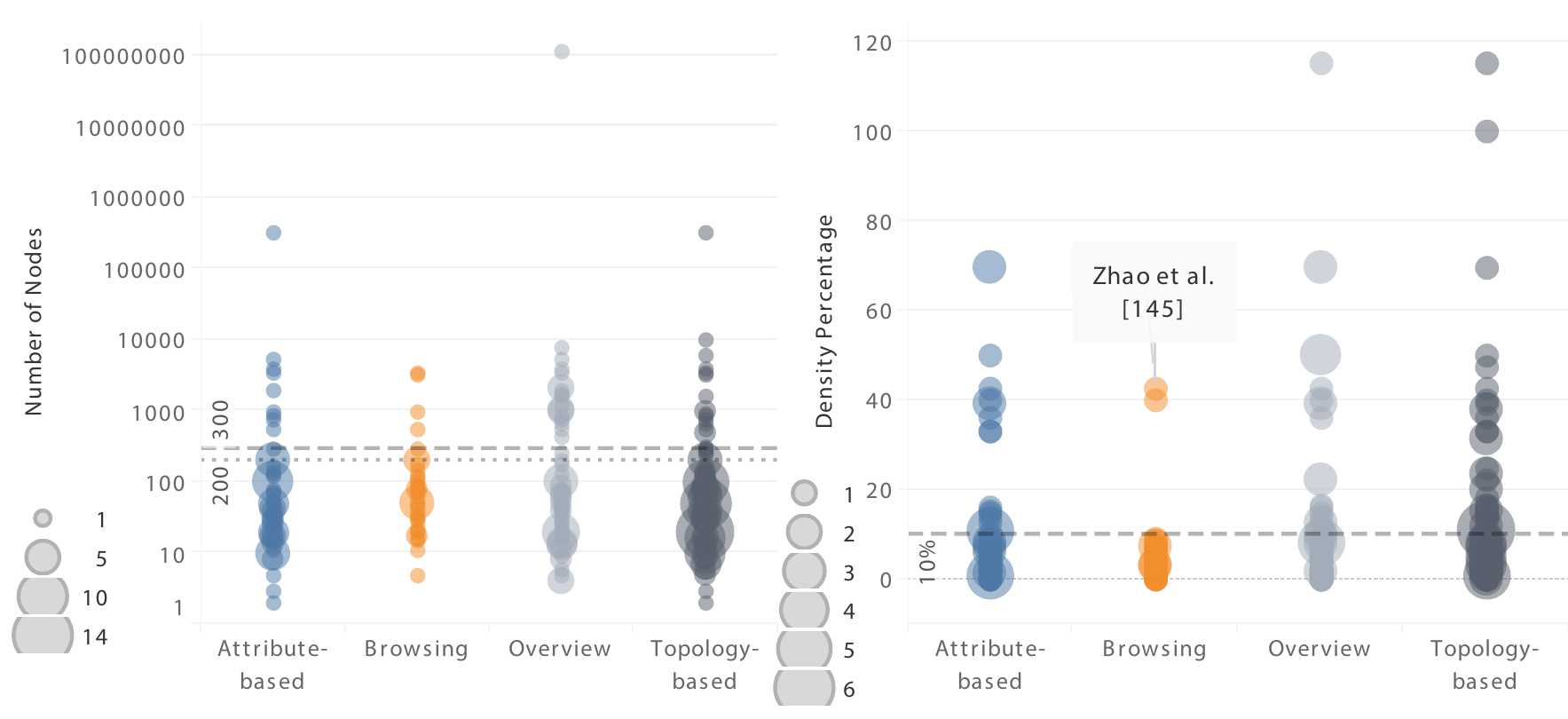}
\caption{\label{fig:tax-tasks} Types of different categories of tasks used in studies with relation to the number of nodes (on the left) and the density (on the right). The areas of the circles reflect the count of studies that use similar values.}
\end{figure*}

In order to understand the combinations of tasks in each study, we plot these as composite nodes in Fig.~\ref{fig:tasks-percentage}. Six studies use only \emph{Topology-based} tasks when using graphs with 500 nodes or more~\cite{PACIFICVIS:5,TVCG:9,CHI:12,EUROVIS:1,TVCG:10}. The first three studies~\cite{PACIFICVIS:5,TVCG:9,CHI:12} use graphs with a simpler structure, i.e.\ trees. Moreover, the trees they use have less than 1,000 nodes. The other three studies~\cite{TVCG:10,CHI:12,TVCG:45}, which use graphs with more than 5,000 nodes, allow participants to interact with the graphs and show only parts of the graph at a given time. %, which are automatically applied based on user gaze movement. 
This implies that tasks and interactions are mutually reinforcing each other in large graph visualisation. 

Marner~\emph{et al.}~\cite{GD:10} use a large graph with 7,885 nodes and 427,406 edges in their study. They project the graph on a wall-sized display and ask participants to untangle it until they achieve a better overview. We believe that even though their process allowed for an \emph{Overview} task, it would have been almost impossible for the participants to perform more complex tasks, such as path-finding or counting triangles. 
Similarly, Kwon~\emph{et al.}~\cite{TVCG:40} use graphs of up to 113 million nodes and 1.8 billion edges in their study, but only ask the participants to rate the similarities between the graphs.
Note that among all tasks, \emph{Topology-based} tasks have been widely used in most of the studies since the connectivity plays a key role for understanding graph structures. Path finding is the most common task (66 over 114 in our survey) among \emph{Topology-based} tasks, and therefore can serve as the primary task for demonstrating the usability of graph visualisation. For example, following the investigation by Bae~and~Watson~\cite{INFOVIS:1}, and Ghoniem~\emph{et al.}~\cite{TVCG:6}, it is reasonable to use node-link diagrams for path-finding purposes when the graph is less than 200 nodes.

Another interesting study using \emph{Topology-based} tasks is conducted by Moskovich~\emph{et al.}~\cite{CHI:4}. They use two graphs with 1,000 nodes. The sparser one has 1,485 edges, while the denser one has 2,488 edges. We believe that they wanted a graph large enough to make the task difficult, especially when finding the immediate neighbours of a given node. The authors propose an interactive navigation technique \emph{Bring-and-Go}, which aids this task by bringing all adjacent nodes closer to a selected node. This study demonstrates the complexity of these tasks on graphs with a large number of edges. Both error rates and performance times increase significantly from the sparser graph to the denser graph.

Some authors discuss the choice of graph sizes in relation to the results of their studies. Huang~\emph{et al.}~\cite{IVJ:7} justify the increase in errors by explaining that human perception and cognitive systems become overburdened when dealing with large graphs, even with 25 nodes and 98 links. Okoe~\emph{et al.}~\cite{EUROVIS:1} use a graph with 900 nodes and 2,500 edges. However, they mention a high error rate of more than $50\%$. They associate this to the difficulty of the tasks. They explain that the large number of edges voided the highlighting advantage of the evaluated technique. Wong~\emph{et al.}~\cite{TVCG:10} mention that they tried to use graphs that were not too complex, but their attempts were not successful for all the tasks. 

Only nine out of the 53 studies that require participants to perform \emph{Attribute-based} tasks, use more than 200 nodes \cite{CHI:2, CHI:6, EUROVIS:9, EUROVIS:11,TVCG:8,TVCG:13,TVCG:38,TVCG:45, GD:6}. Two of these studies aggregate multiple nodes into singular representations: \emph{motifs}~\cite{CHI:2} and \emph{metanodes}~\cite{EUROVIS:9}. It is natural to assume that \emph{Attribute-based} tasks would be difficult on graphs with too many elements, since \emph{Attribute-based} tasks are related to data attributes associated to nodes and links. While this is backed by our survey for number of nodes, the number of edges ranges between 10 and 1,000 for most studies using \emph{Attribute-based} tasks.

For \emph{Browsing} tasks, we assumed that experimenters would use sparse graphs. This is backed by our findings as shown in Fig.~\ref{fig:tax-tasks}. Also, in addition to the density of the graphs, studies that require performing \emph{Browsing} tasks use graphs with few number of nodes ($\leq 200$), with the exception of five outliers \cite{CHI:4, EUROVIS:9, IVJ:10, TVCG:25,TVCG:38}. Four of these outliers use interactive highlighting to assist the participants in performing the task \cite{CHI:4, IVJ:10, TVCG:25,TVCG:38}, while the fifth~\cite{EUROVIS:9} uses \emph{metanodes} and \emph{metaedges}, which are aggregations of multiple nodes and edges into singular representations. 
As seen in Fig.~\ref{fig:tasks-percentage}, \emph{Browsing} tasks are only used on very sparse graphs ($<10\%$), except for one study by Zhao~\emph{et al.}~\cite{CHI:7} that uses graphs with 40\% and 42.9\% densities. The latter allows participants to highlight specific nodes and their connections, thus showing small subsets of edges at a given time. %Since \emph{Browsing} tasks require understanding the connections of a network, an increase in graph density will naturally increase cognitive load. 
%\emph{Attribute-based} tasks require an understanding of the attributes of both the objects and connections of the network. Thus,  

In summary, \emph{Topology-based} tasks are the most common tasks to evaluate graph visualisation; nonetheless, the number of \emph{Topology-based} tasks is relatively reduced when using graphs with more than 300 nodes. In such cases, \emph{Overview} tasks become more common, especially when the tasks are assisted with multiple interaction techniques. Sparse graphs ($<10\%$) with few nodes ($\leq 200$) are used when asking the participants to perform \emph{Browsing} tasks. Similarly, \emph{Attribute-based} tasks are not common in studies that use graphs with more than 200 nodes.

\subsubsection {Interaction}
\label{sec:interaction}

\begin{sidewaysfigure*}
\centering
\includegraphics[width=\linewidth]{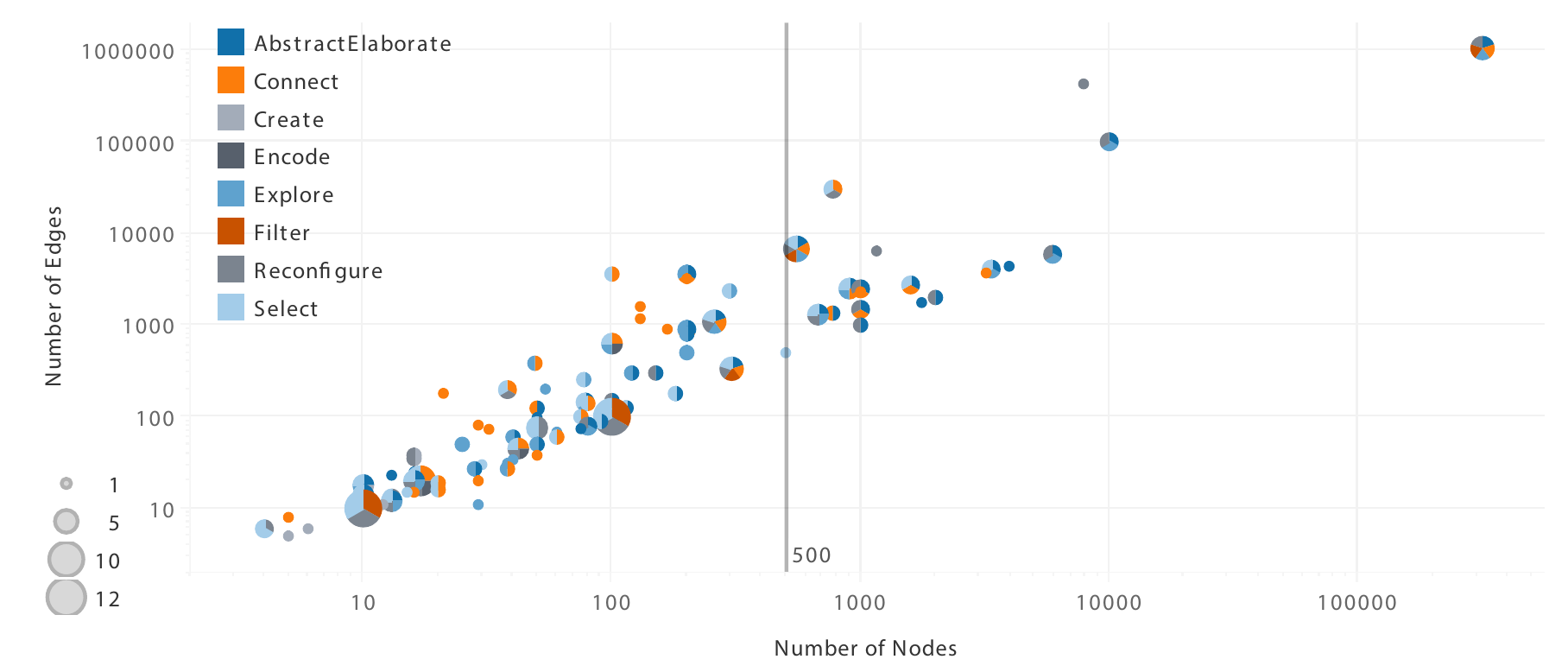}
\caption{\label{fig:interaction-nodes-edges} Types of different interaction used in studies with relation to the number of nodes and the number of edges. Both axes have log scales.}
\end{sidewaysfigure*}

Early graph experiments did not make much use of interaction; their focus being on the interpretation of static graph drawing--often being simply presented on paper (e.g.\ \cite{GD:20,GD:17,GD:18}). In more recent years, several experiments have tested the worth of node-link diagrams using interactive systems that permit more extensive exploration of the relational information. In such cases, the use of interaction techniques is often a crucial component of the study design.

We used the interaction taxonomy of Yi~\emph{et al.}~\cite{yi2007toward} as a means of classifying the different types of interaction used in the experimental studies: \emph{Select} allows users to mark objects on screen as interesting; \emph{Explore} enables users to navigate a hidden subset of data; \emph{Reconfigure} changes spatial arrangement; \emph{Encode} allows users to transform data values to their preferred visual language (e.g.\ colour, size, or shapes); \emph{Abstract/Elaborate} enables users to adjust the level of detail of a data representation; \emph{Filter} allows users to prevent the display of data fulfilling given conditions, and; \emph{Connect} allows the highlighting of relationships between data. For the purposes of this survey, we added an additional category: \emph{Create}, which allows users to generate graph drawings.

Of the 152 studies, 80 used no interaction at all. Table~\ref{tab:interaction} shows how the remaining 72 used interaction (noting that several studies used more than one interaction technique). Of all the occasions when interaction techniques were used, \emph{Abstract/Elaborate} was most common (20\%%43\% of the surveyed studies that use any interaction
), followed by \emph{Select} (19\%%33\%
), \emph{Explore} (16\%%38\%
) and \emph{Reconfigure} (16\%%32\%
). Thus, there is no single category that dominates (unlike our finding with the task category).

Fig.~\ref{fig:interaction-nodes-edges} shows the different types of interaction used in the studies, with respect to graph size. We note that 25 of 60 studies (42\%) use a single interaction type for those graphs with 500 nodes or fewer, while for those studies using graphs with more than 500 nodes, this ratio drops to 32\% (6 out of 19 studies).

\begin{figure*}
\centering
\includegraphics[width=\linewidth]{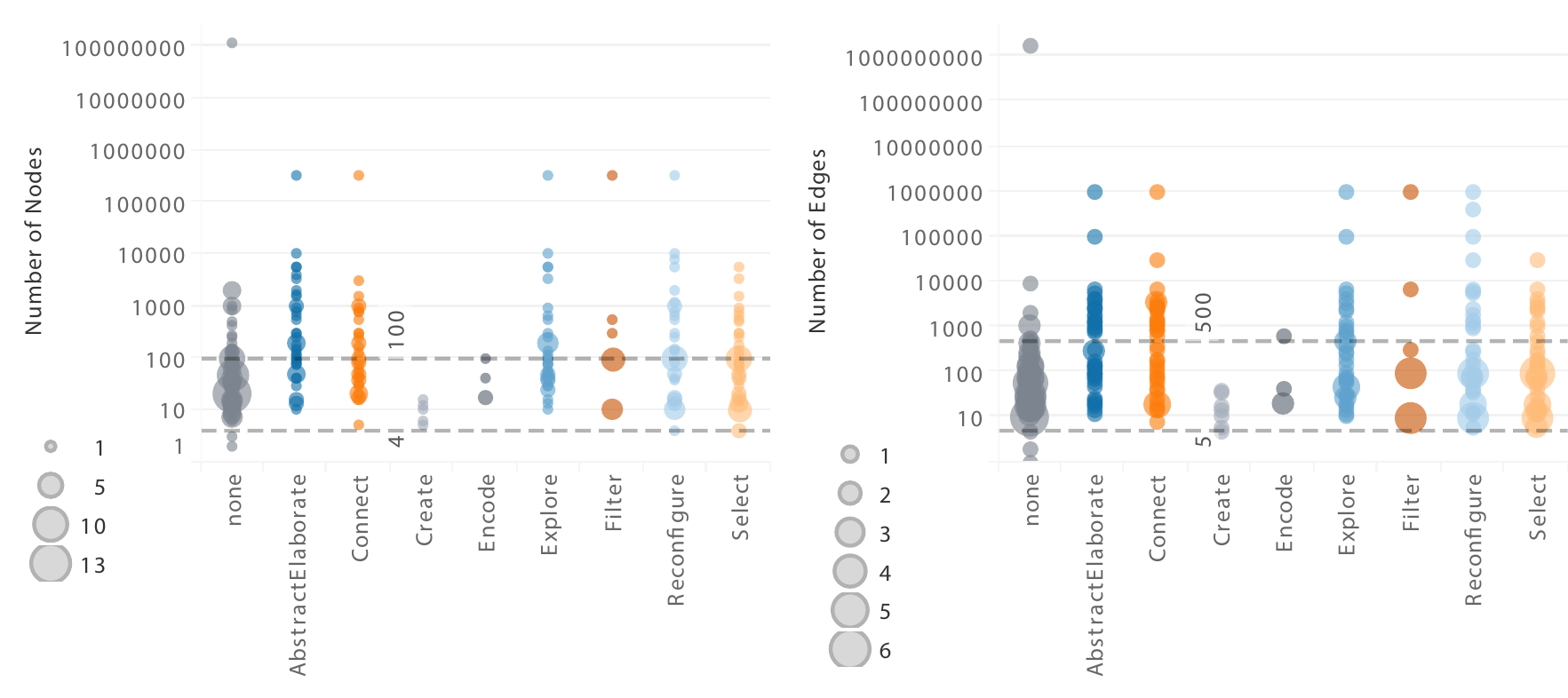}
\caption{\label{fig:interaction-nodes-edges-density} Interaction types used in studies with relation to the number of nodes (on the left), and the number of edges (on the right).}
\end{figure*}

Tiny graphs (e.g.\ 4 nodes and 5 edges) are easy to understand without the support of interaction (see lower dashed lines shown in Fig.~\ref{fig:interaction-nodes-edges-density}), although even studies that used graphs with as few as ten nodes and ten edges used interaction. The majority of studies that do not use any interaction use graphs with 100 nodes or less (70 / 80 studies, 88\%) and 500 edges or less (71 / 80 studies, 89\%) (upper dashed lines in Fig.~\ref{fig:interaction-nodes-edges-density}). These bounds allow us to group the interaction techniques into two sets: techniques used on graphs with greater than 100 nodes and 500 edges: {\emph{Abstract/Elaborate}, \emph{Connect}, \emph{Explore}, \emph{Reconfigure}, \emph{Select}}; and techniques used on graphs with fewer than 100 nodes and 500 edges: {\emph{Encode}, \emph{Create}, and \emph{Filter}}. If we consider the degree of effort required by the user in applying these techniques, the first set can be considered relatively straight-forward -- the techniques are usually tested by having participants apply the built-in functionality to look at the data in a different way until the answer is obvious. By contrast, \emph{Encode} and \emph{Create} require more effort on the part of the users, since (possibly creative) decisions need to be made. 

Interaction tasks can come with a time trade-off, however, for example, finding labels inside an abstract node (which represents a set of aggregated nodes), could be more time consuming due to the interaction, than the task of finding labels outside of it, which is always completed faster \cite{CHI:2}. Care, therefore, needs to be taken in the interpretation of the timing data collected (especially if participants are allowed unlimited time to perform their task), since such data might include superfluous interaction.

To conclude, the combination of multiple interactions can improve the visualisation of large graphs, and this is more significant as graph size increases. Based on the studies we analysed, the floor effect of interaction lies at the boundary of nodes equal to 4 and edges equal to 5, with ceiling effect at 100 nodes and 500 edges.

\begin{table}
\caption {The type of interaction used within the studies.}\label{tab:interaction}
\centering
{\scriptsize
	\begin{tabu} to \columnwidth {cXc}%
	\toprule
    \textbf{Interaction} & \textbf{References} & \textbf{\# of Studies}\\
	\midrule
	None & \cite{CHI:3,CHI:5,CHI:9,CHI:11,APVIS:1,CGF:1,DIAGRAMS:1,DIAGRAMS:2,DIAGRAMS:4,EUROVIS:4,EUROVIS:7,GD:1,GD:2,GD:3,GD:5,GD:7,GD:9,GD:11,GD:12,GD:13,GD:15,GD:16,GD:17,GD:18,GD:19,GD:20,IVJ:1,IVJ:2,IVJ:4,IVJ:5,IVJ:6,IVJ:7,IVJ:8,PACIFICVIS:1,PACIFICVIS:2,PACIFICVIS:3,PACIFICVIS:4,PACIFICVIS:5,PACIFICVIS:6,TVCG:4,TVCG:12,TVCG:13,TVCG:19,TVCG:20,TVCG:22,TVCG:23,TVCG:26,TVCG:28,TVCG:32,TVCG:33,VINCI:1,VLHCC:1,TVCG:34,TVCG:36,TVCG:37,TVCG:39,TVCG:40,TVCG:41,TVCG:44,IVJ:12,IVJ:13,GD:23,CGF:2,EUROVIS:10} & 80 \\
Abstract-Elaborate & \cite{CHI:2,CHI:3,CHI:4,CHI:8,CHI:10,CHI:12,DIAGRAMS:5,EUROVIS:1,EUROVIS:3,EUROVIS:8,EUROVIS:9,GD:4,GD:6,TVCG:2,TVCG:5,INFOVIS:2,INFOVIS:3,TVCG:8,TVCG:10,TVCG:14,TVCG:16,TVCG:18,TVCG:21,TVCG:24,TVCG:29,TVCG:30,TVCG:38,TVCG:45,IVJ:11,GD:22,EUROVIS:11} & 32\\
Connect & \cite{CHI:1,CHI:4,CHI:6,CHI:7,CHI:10,DIAGRAMS:5,EUROVIS:1,GD:8,IVJ:9,IVJ:10,TVCG:6,TVCG:14,TVCG:21,TVCG:27,TVCG:31,CHI:13,TVCG:35,TVCG:38,TVCG:42,TVCG:43,TVCG:45,IVJ:11,IVJ:14,GD:21,GD:22,EUROVIS:11} & 29 \\
Create & \cite{DIAGRAMS:3,TVCG:2,TVCG:15,TVCG:17} & 4 \\
Encode & \cite{CHI:10,TVCG:27,TVCG:43} & 4 \\
Explore & \cite{CHI:10,CHI:12,EUROVIS:1,EUROVIS:2,EUROVIS:3,EUROVIS:5,EUROVIS:6,EUROVIS:8,EUROVIS:9,GD:4,GD:8,TVCG:1,TVCG:2,TVCG:7,INFOVIS:2,TVCG:8,TVCG:10,TVCG:16,TVCG:21,TVCG:24,TVCG:25,TVCG:27,TVCG:38,TVCG:45,GD:22} & 26 \\
Filter & \cite{CHI:10,IVJ:3,TVCG:11,TVCG:38,TVCG:45,EUROVIS:11} & 9 \\
Reconfigure & \cite{CHI:3,CHI:4,CHI:6,CHI:10,CHI:12,GD:6,GD:10,IVJ:3,TVCG:3,INFOVIS:2,TVCG:10,TVCG:11,TVCG:12,TVCG:17,TVCG:24,TVCG:38,TVCG:43,TVCG:45,GD:22,EUROVIS:11} & 25 \\
Select & \cite{CHI:6,EUROVIS:1,EUROVIS:9,GD:14,IVJ:3,IVJ:9,TVCG:2,TVCG:3,TVCG:5,INFOVIS:2,TVCG:8,TVCG:9,TVCG:12,TVCG:18,TVCG:24,TVCG:25,TVCG:27,TVCG:35,TVCG:38,TVCG:43,IVJ:11,GD:21,GD:22,EUROVIS:11} & 30\\
 \bottomrule
\end{tabu}
}
\end{table}

\subsubsection {Application Areas}

There are several domain-specific studies where the aim is enhanced understanding of the content, and so the tasks were clearly focused on the domain knowledge (e.g.\ \cite{IVJ:3, DIAGRAMS:3}). For example, Tanahashi~\emph{et al.}~\cite{CGF:1} performed a comparative study of four different ways of presenting data for the purposes of introducing information visualisation to novices, where a graph drawing was one of the visualisation types. Their analysis not only considered the efficacy of the visualisations, but also looked at two different types of learning (active and passive), and two different teaching methods (top-down and bottom-up). They were therefore able to propose guidelines for writing effective information visualisation tutorials. North~\emph{et al.}~\cite{IVJ:1} compared two different types of evaluation (benchmark and insight) using three different visualisation alternatives depicting gene expression data, and all tasks were related to understanding of the data.

While some application areas have well defined and restricted characteristics for the networks under investigation, in others there can be a large range of impacting factors that potentially affect scalability. In the life sciences, a variety of network types are investigated, ranging in size from a few dozen to a few million nodes, and showing a similar variety in other network characteristics, e.g. density or diameter. In addition, the required tasks can differ significantly between use cases, e.g. from deciding reachability to the detection of dynamic patterns, which affects the limits of readability. RNA sequence graphs, like the ones used in~\cite{GD:10}, can have up to several million nodes and edges, and dense local structures, but they often have a very sparse global structure, making the visual detection of so-called repeats (loops that indicate repetitive structures in RNA sequences), a feasible task. On the other hand, metabolic pathways like the ones used in~\cite{GD:6} are often planar or near-planar graphs with low local and global density, but require that the semantics associated with the metabolic flow are incorporated in the layout and the visual representation. While these pathways are parts of a large and complex network of metabolic reactions in an organism, the visual analysis is often restricted to such sparse sub-networks that have a specific functionality, e.g. the synthesis of a particular biomolecule. 

%\end{itemize}

\subsection {Graph Drawing Factors}
\label{sec:gdFactors}
%{\bf Graph Drawing factors intro goes here}
For several decades now, the field of Graph Drawing has led to the development of efficient algorithms for layout computation as well as graph visualisation metaphors, and has also investigated the impact of the resulting visualisations on readability and task performance. % while a focus is on computational complexity
For some years, a main focus has been on the computational complexity and scalability of algorithms, but since the development of methods that scale to several million nodes and edges, the focus has shifted to the visual complexity and human interpretability of the resulting layouts~\cite{Eades17}.

As layout methods differ in computational scalability, in their performance on certain graph classes, and also in the features and characteristics of the resulting layouts, the interplay between graph structure and layout method used in studies will strongly impact the limits of cognitive scalability. 
Comparisons between different methods are rare, as the selection is often motivated by real-world application requirements. While constraint-based methods can create high-quality layouts, which might be of interest for studying the limits of cognitive scalability, the methods do not scale well regarding the computational complexity. Thus, for large graph sizes, researchers have to resort to fast heuristics, e.g. multi-level force-based methods.
%The structure of the graphs used in studies, or more general the graph classes, can be motivated by real-world application requirements, 

\subsubsection {Graph Type and Structure}

%\todo{SD: graph types: 86 undirected, but only 23 directed, 17 tree}
%%KK I do not see those numbers from the study sheet

While the number of nodes and the density can give a first indication on the complexity of a graph with respect to cognitive scalability, a more fine-grained description of the structure is necessary to investigate its impact. Certain graph classes might result in layouts with minor quality, e.g. low diameter graphs like friendship networks can lead to hairball drawings when standard force-directed methods are applied. On the other hand, graphs with a globally sparse structure like the ones used by Marner~\emph{et al.}~\cite{GD:10} are well suited for untangling and structure identification tasks, even when the number of nodes is huge, as the global and local structures simply scale with the size.

A further distinction can be made regarding the use of directed and undirected graphs. While 103 of the studies used undirected graphs, only 33 of them used directed graphs, and 15 used trees. With the exception of the two near-tree sparse graphs from the \emph{OnGrax} study~\cite{GD:6}, and one graph from a study on directed edge representations~\cite{PACIFICVIS:2}, all graphs with more than $200$ nodes are either undirected or trees.
Some studies based their graph selection on real world examples. In some cases, the graphs were taken from specific application areas. Most commonly: social networks \cite{CHI:3,CHI:6,CHI:10, GD:1,GD:8,GD:12,GD:16,IVJ:5,IVJ:7,IVJ:8}, followed by co-authorship networks \cite{CHI:7,TVCG:11, TVCG:19, TVCG:24,TVCG:31} and biological networks \cite{CHI:12, GD:6, GD:10, IVJ:1, TVCG:12}.
\emph{Shi et al.}~\cite{TVCG:24} use four dynamic graphs from two domains--communication and co-authorship networks.

Other studies used extractions of real graphs, which were closely related to ones in practice. For example, North~\emph{et al.}~\cite{IVJ:1} used graphs that represented a subset of a biological data set, while others generate graphs with similar size to graphs that commonly appear in usage. 
Tan~\emph{et al.}~\cite{TVCG:28} use two trees with 127 nodes with the number of nodes chosen to be typical of tournament brackets considered in fantasy football leagues.

\subsubsection {Layout Method}

The distribution of layout methods in the investigated studies shows the expected dominance of force-directed methods. Variants of this class of methods scale well computationally, and appear to be the preferred method used in a variety of publicly available graph visualisations and systems. Together, with the linear time tree layout methods, these methods are the only ones used for studies with more than one thousand nodes.
More than 50\% of the studies used a force-directed layout to draw the networks, followed by multiple types, indicated in around 18\% of the studies. \deleted[id=HW]{Even though only a few studies used orthogonal layouts, it is worth mentioning that a study on human-like layout investigated an orthogonal layout style \cite{TVCG:12}.}
% Is that due to the challenge, or the gap?
Note that our classification of layout methods specifies the initial layout the subjects are presented with, and does not consider whether the subjects could manually, or by means of an algorithm, change the layout in an interactive interface.

%\todo{HP - not sure this paragraph fits - it is not something we have discussed. KK - it is not a finding, just a general observation}
While the practical computational scalability of many methods changed due to improved algorithms and implementations, the relative performance stayed the same with force-directed and tree layouts being by far the fastest, and methods that require solvers, e.g. constraint-based methods, being slowest.

Borkin~\emph{et al.}~\cite{TVCG:8} present a radial-based tree layout to display file system provenance and motivate their choice, with the failure of node-link diagrams to show a high-level summary for the large-scale provenance data graphs. Their study results indicate that users were more efficient with the interactive radial layout representation than with an existing conventional node-link diagram tool.  

The range of the basic graph size metrics for each of the layout methods fits the expectation. Manual layouts are only performed on graphs of $120$ nodes or less, with the notable exception of the study on the collaborative graph visualisation system \emph{OnGrax}~\cite{GD:6}. In this system, however, an initial layout was given, which was manually created by domain experts, and the user could simply rearrange this layout manually.

\subsection{General Study Design Factors}
\label {sec:studyDesign}

Just over half the studies (92 / 152 studies, 61\%) follow a typical design of asking participants to perform graph reading tasks under different conditions (the independent variables - often different layouts, different visualisations or different interaction techniques) and collecting response time and accuracy data as the dependent variables (with some also collecting preference choices). Most studies rely on these three dependent variables to measure the efficiency of the visualisations at hand. 

With respect to graph size, there is no discernible difference in the sizes used for typical graph reading studies and the others--both categories have similar distribution of graph sizes.

Some experiments collected process data as the main dependent variable (e.g.\ \cite{TVCG:9, PACIFICVIS:5}). 
Others collected eye-tracking data \cite{TVCG:41,CHI:13,TVCG:5,DIAGRAMS:5}.
\emph{Wong et al.}~\cite{INFOVIS:2} counted the number of mouse actions (click, move, zoom, pan, and turn) as a measure of extent of interaction with different forms of visualisation of labelled graphs, while \emph{Nekrasovski et al.}~\cite{CHI:12} looked particularly at the mouse drag action.

There were 129 studies that used a within-participants study design, 22 used between-participants, and one used a mixture of both. Within-participants studies are popular because they have the advantage of eliminating any effects relating to variability between the participants, and, while they may be subject to the learning effect, the effects of this are easily mitigated by appropriate randomisation. They do, however, tend to take longer than between-participants experiments, where the participants can have more time to work with only a selected few of the stimuli (rather than all of them, as it is in the within-participants case). Thus, the tasks for within-participants studies tend to be smaller and simpler than those used in between-participants studies.

We might expect that there would be more between-participants' experiments in recent years, since such experiments are more suitable for crowd-sourcing: within-participant experiments tend to take too long to be appropriate for crowd-sourced participants. However, this is not the case – there is a similar publication year profile for both categories of study.

%% file: sections/size-rationale.tex
%!TEX root = ../main.tex 
\section{Size Rationale}
\label{sec:size-rationale}
In the previous sections we discussed how different metrics could affect scalability. This was mainly done based on what we could gather from the aim and the results of the studies. In Section~\ref{sec:basic-metrics}, we reviewed the range of number of nodes and edges for graphs used in studies, and discovered that most studies using large numbers of nodes and edges want to highlight the advantages of using specific techniques.
However, in some cases, the authors explicitly mention the reasons for picking a specific number of nodes or edges. Some authors performed pilot studies which allowed them to determine the size of graphs that would best suit their evaluation. Others were based on the authors' experiences and understandings of the requirements. We discuss these further in the following sections.

\subsection{Pilot Studies}
\label{sec:pilot}

The papers in our survey rarely mention pilot studies that determine a ceiling or floor affect to scalability. We believe that many conduct pilot studies but do not mention them explicitly. In this section, we discuss a select few that provide the reasoning for their pilot studies and the decisions made with regards to factors that affect cognitive scalability.
 
Archambault~\emph{et al.}~\cite{EUROVIS:9} mention that the largest graph size for their study was determined by pilot studies. They start with small, medium and large sized graphs; however the pilot participants could not complete the tasks on the large graphs. Thus, instead of the large graph, they use a smaller graph. The largest graph they use has 3,351 nodes and 4,083 edges. Archambault~and~Purchase~\cite{IVJ:8} use a pilot study to find a reasonable graph size. They use a maximum of 50 nodes and 100 edges. Similarly, Dawson~\emph{et al.}~\cite{IVJ:2} mention a pilot study to balance density and difficulty. They use graphs with 75 nodes and 150 edges. 

Some design their pilots to find reasonable graph densities to allow the participants to finish the tasks in a restricted time limit.
Kobourov~\emph{et al.}~\cite{GD:4} find 120 nodes and 2.5 density as maximum measures to complete the tasks under two minutes, while others use pilot studies in order to explore different layouts and graph structures for specific tasks; e.g.\ Netzel~\emph{et al.}~\cite{TVCG:5} aim at finding thresholds for assisting tasks of finding the longest link and biggest cluster.

Saket~\emph{et al.}~\cite{TVCG:16} mention a pilot study where they chose 50 nodes as minimum and 200 nodes as maximum. They also chose three density levels $N$, $2N$, and $4N$.
Borkin~\emph{et al.}~\cite{TVCG:8} conducted a pilot study to determine the boundaries between the easy and difficult tasks. They mention that graphs of 10s, 100s, 1000s, and 10,000s of nodes were compared. They classified trees with 42 to 346 nodes as easy, while trees with 1,192 to 5,480 nodes as hard. 
Similarly, Marriott~\emph{et al.}~\cite{TVCG:15} note that their pilot study showed that larger graphs beyond 6 nodes were too difficult to be memorised. 
Conversely, Xu~\emph{et al.}~\cite{TVCG:4} rely on a pilot study to choose 100 as the maximum number of nodes for their graphs. They further increase this to 200 in their second study, due to the lack of a ceiling effect in their first study. 

\subsection{What is Small? What is Large?}

Testing the boundaries of readability/scalability may generally not be a primary goal for experimental studies. Instead, researchers might pick a size range that they consider acceptable in order not to have scalability as a confounding factor in their results. In addition, technical limitations, (e.g.\ screen size and resolution, and also typical requirements from application areas, like characteristics of occurring networks of interest), might play an important role in the determination of graph sizes, but are often not reported explicitly and will also change over the years. Furthermore, there might be standard benchmark sets used, or simply graphs picked based on availability, instead of using graphs that allow one to investigate scalability effects.

Some studies do not mention pilots, but justify their choices of graph size as an attempt to meet a particular requirement for their studies.
Archambault~\emph{et al.}~\cite{TVCG:1} use real graphs, with the largest having 60 nodes and 68 edges. They explain that they chose two data sets each consisting of graphs with realistic size and structure. 

Sometimes, the rationale is task-oriented. For example, Kieffer~\emph{et al.}~\cite{TVCG:12} justify their use of small graphs to allow the participants to manually draw the graphs in a reasonable amount of time. Similarly, Purchase~\emph{et al.}~\cite{TVCG:2} use two graphs with 10 nodes and 11 and 18 edges, to make it manageable for the task of drawing the networks. Blythe~\emph{et al.}~\cite{GD:16} note that they use a small graph in order not to overwhelm the participants with the amount of information. They use a small graph with 12 nodes and 24 edges. 

Others, such as Hlawatsch~\emph{et al.}~\cite{TVCG:22}, justify using small graphs to avoid the need for interaction. They use ten graphs with eight nodes and 22 to 40 edges. They also use ten graphs with 20 nodes and 147 to 264 edges. Similarly, Alper~\emph{et al.}~\cite{TVCG:19} chose not to vary the graph size drastically in order to avoid the requirement of zooming.

Zhao~\emph{et al.}~\cite{CHI:1} justify their size selection to fit the diagrams to screen. They use graphs with 167 nodes and 902 edges in their evaluation of a visualisation called \emph{MatrixWave}. Kadaba~\emph{et al.}~\cite{TVCG:33} mention that they needed graphs that were small enough to be memorisable. They use a daisy-structured graph with 11 nodes. In a second study they use smaller graphs with 3 nodes and 2 edges.
Holden~and~Van~Wijk~\cite{CHI:9} explain that they wanted to generate graphs with an adequate number of vertices, without causing a large amount of visual clutter caused by an excessively high edge density.

In contrast, some studies use large graphs in order to highlight the benefits and improvements of some techniques with respect to scalability.
Lee~\emph{et al.}~\cite{TVCG:21} justify their selection of graphs with 200 nodes as complex graphs. They also state that 200 nodes is considered to be an upper bound for currently studied food webs.
Huang~\emph{et al.}~\cite{PACIFICVIS:6} mention that they choose graphs with a density ranging from 10\% to 20\%, and their drawings have the same crossing ratio of 40\% in order to have reasonably complex graphs. The largest graph they use has 50 nodes and 245 edges.
Dwyer~\emph{et al.}~\cite{TVCG:3} state that the graphs they used were larger in size (50 nodes), than ones (17 nodes) used in another study which inspired their work, while Okoe~\emph{et al.}~\cite{GD:22} justify their selection of a graph with 258 nodes and 1090 edges as larger than previously used graphs, yet sufficiently small to be evaluated in a browser.

To conclude, some authors provide a rationale for their selection of graphs with a specific range of node and edge counts. These often resemble our discussions and hypotheses, nonetheless, we wanted to keep a clear separation between what constitutes our opinions and the rationale provided by the experimenters.

%% file: sections/history.tex
%!TEX root = ../main.tex 
\section{Research Trends}
\label{sec:history}

By analysing our survey data from a historical perspective, we tried to gain some insights about the development of the research community, including trends with respect to graph sizes, participants, tasks, and interaction types used in the studies.
%We also briefly discuss trends related to publication venues.

%\subsection*{Venues}
%% based on Tableau "perc_venue"
Of the studies surveyed in this paper, 40\% 
were published in TVCG/InfoVis, 16\% at Graph Drawing, 13\% in the Infromation Visualisation Journal, and another 11\% at CHI.  
%% based on Tableau "numberofpapersacrossdiff_venuesandyears"
Historically, the first seven studies~\cite{GD:16,GD:18,GD:20,GD:3,GD:17} were all published at Graph Drawing between 1995 and 2000. The first study~\cite{IVJ:3} in the Information Visualisation Journal appeared in 2002. The first one~\cite{INFOVIS:3} in TVCG/InfoVis was in 2003, whereas the first one~\cite{CHI:12} at CHI in 2006, and the first one~\cite{EUROVIS:7} at EuroVis in 2009. 

\begin{sidewaysfigure*}
\centering
\includegraphics[width=\linewidth]{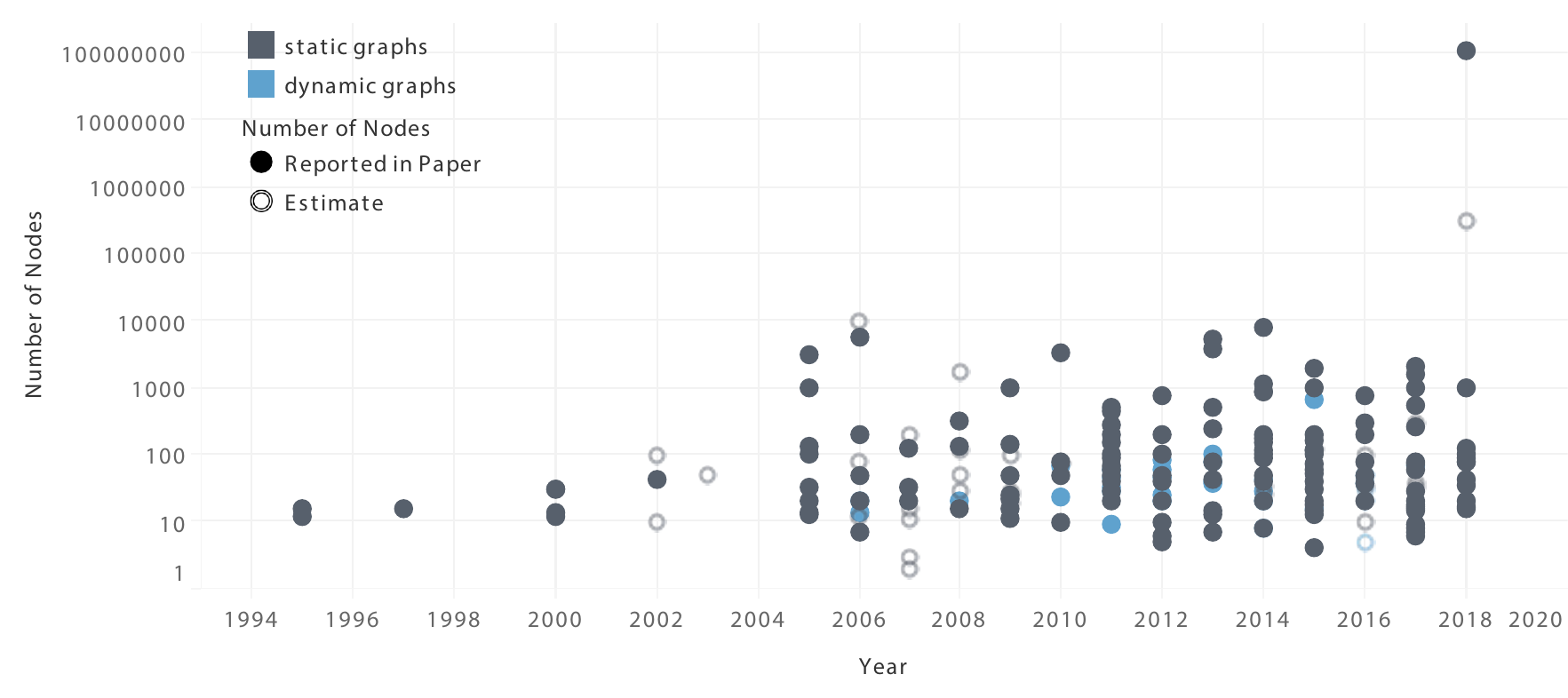}
\caption{\label{fig:nodes-years-dynamic}The number of nodes of static versus dynamic graphs used in studies for the period from 1995 to 2016.}
\end{sidewaysfigure*}

The seminal papers~\cite{GD:16,GD:18} of 1995 focused on static graphs and were followed by another ten papers (with a total of 16 studies) on static graphs. The first study on dynamic graphs~\cite{GD:11} was published in 2006, more than ten years after the first one on static graphs. 
As we cover a period of 24 years from 1995 to 2018 in this paper, it is reasonable to compare the first and second half of the studied period, with the first half ending in 2006, and the second half starting in 2007. Whilst during the first half, the average publication frequency was around two studies per year, it considerably increased to 11 studies per year for the second half. Of these studies, on average, 9 focussed on static and 2 on dynamic graphs.

Fig.~\ref{fig:nodes-years-dynamic} shows the number of nodes of static and dynamic graphs used in studies in different years. While for static graphs the number of nodes increased from below 20 in 1995 to several thousand ten years later, the graph size of dynamic graphs stayed 100 or below (with one exception~\cite{TVCG:24} in 2015).

Since we observed a considerable increase in the graph size for static graphs, we expected that researchers would also try to recruit more participants for their studies. It turns out that there was actually a big difference, if we compare the first and the second half of our studied period. In the first half the median number of participants, for studies that used static graphs, was 14. In the second half it increased to 21\deleted[id=VY]{(average 44\%)}. Interestingly, a  closer look at the first half of the studied period reveals a dramatic drop of the median number of participants. It decreased from 75\deleted[id=VY]{(average 63\%)} for the original 7 studies~\cite{GD:16,GD:18,GD:20,GD:3,GD:17} (all published at GD)
 of the period from 1995 to 2000 to  9 \deleted[id=VY]{(average 18\%)} for the subsequent 11 studies~\cite{IVJ:3,IVJ:4,INFOVIS:3,GD:12,IVJ:9,IVJ:10,INFOVIS:2} from 2001 to 2005 (mostly published in the Information Visualisation Journal).

%% file: sections/conclusion.tex
%!TEX root = ../main.tex 
\section{Summaries of Findings and Discussion}

\subsection {Author Personal Experiences}

Several of the authors of this paper have conducted experimental studies included in this survey; we therefore have our own insights into the problem of choosing an appropriately-sized graph for an empirical study. In many cases (e.g.~\cite{DIAGRAMS:2, GD:17, GD:18,EUROVIS:9}), the original graph chosen was found to be too large during piloting, and scaled down. This was not because it was thought that the participants would not be able to complete the task correctly, but simply because it was necessary to keep the tasks short in duration so that the experiment did not take too long, especially when using a within-participants’ study design. Even in cases where real-world data was used (e.g.~\cite{IVJ:8, PACIFICVIS:4}), these graphs were filtered to make them of an appropriate manageable size for within-participants experiments. On the one occasion when a complex UML diagram was used in a between-participant study, the variability between subjects was so great that no sensible results were achieved: one of the many times when data was simply discarded. 

Another common way we selected our graph sizes (and, in some cases, graph structure) was with reference to published work, particularly if the study was deliberately building on prior experiments -- even our own. The size and structure of the graphs used in \cite{TVCG:3,TVCG:2} were based on those used by \emph{van Ham and Rogowitz}~\cite{TVCG:17}; the number of slices in the dynamic graph used in \emph{Archambault and Purchase}~\cite{GD:13} was based on the number used in our own prior work~\cite{TVCG:1}. Evidence that the choice of size of graph has proven successful in prior work is a clear pointer as to an appropriate size to choose for experiments that are similar in scope.  

A further way to define graph characteristics for experimental studies is based on the requirements from application areas. In~\cite{GD:10}, 
the graphs were provided by biologists that used the graphs in a previous publication to analyse sequencing data by applying network analysis and visualisation. The graph set was chosen to exploit and test the affordances of a wall-sized display.

\subsection{Recommendations for Experimenters}

Here we present our recommendations, based on our findings, for future design and reporting of studies.
	
We recommend that authors of study papers report precisely the number of nodes and edges used in studies, as well as the number of timeslices for dynamic graphs.  In the case of interactivity, such as semantic zooming or neighbourhood browsing, the number of nodes visible on screen should also be clearly recorded to differentiate between scalability due to interaction versus cognitive and perceptual scalability. Similarly, studies that use aggregation to collapse parts of the network into less numerous glyphs, should report on the number of glyphs visible. 

Authors of papers describing studies on network analysis tasks should explicitly state how they controlled the size and density of the network data tested.  This need not entail extra work.  Simple recording of observations during piloting could be greatly informative to the kind of meta-analysis we have performed in this survey. Alternatively, if the choice of size is arbitrary or based on previous studies, explicitly stating the provenance of the choice can help readers to better reason about size effects.

We would call for studies that explicitly test scalability of network visualisation by testing more than a couple of different graph sizes to search for ceiling and floor effects and controlling for other variables. Moreover, we call for the reporting of whatever ceiling and floor effects that are found.

\section{Conclusions}

\subsection{Specific Findings}

We report the following key findings:
\begin {itemize}[leftmargin=*, itemindent=1em, itemsep=1mm]
	\item There are some clear `default' numbers of nodes and edges used by most studies. At a maximum, most studies use graphs with 100 nodes, followed by 50 nodes and 100 edges, while their smallest graphs have 20, followed by 50 and 10 nodes, and 10 edges.  The round numbers used by most studies suggest the choice of size to test is somewhat arbitrary, as opposed to being based on empirical evidence of cognitive limitations.
	\item 80\% of studies (121 / 152 studies) use graphs with 100 nodes or less,  while only 37\% (56 / 152 studies) use graphs with more than 100 nodes.
	\item 74\% of studies (113 / 152 studies) use graphs with 200 edges or less, while only 34\% (52 / 152 studies) use graphs with more than 200 edges.
	\item 70\% of studies (23 / 33 studies) that use graphs with more than 200 nodes use interaction and aggregation techniques to show only parts of the network to the participants at a given time.
	\item Most of the studies that use graphs with more than 1,000 edges evaluate tools that are intended to be able to cope with a substantially large number of edges, or are oriented towards performing well for networks of specific structure, e.g.\ densely connected communities.
	\item Only $12\%$ (18 / 152 studies) of the studies surveyed use graphs with a density of more than $20\%$.
	\item Studies that use graphs with more than 20\% density either use small graphs ($<10$ nodes), evaluate matrix representations, or evaluate edge bundling and compression techniques.
	\item 32\% of studies (7 / 22 studies) that use dynamic graphs use six timeslices. This is often to best fit small multiples representations to screen.
	\item The most common type of tasks is \emph{Topology-based} (114 / 152 studies, 75\%), however for graphs with 1,000 nodes or more, \emph{Overview} tasks prevail (13 / 18 studies, 72\%).
	\item \emph{Attribute-based} tasks are commonly used for graphs with 200 nodes or less (47 / 53 studies, 89\%) and less than 10\% density (42 / 53 studies, 79\%).
	\item \emph{Browsing} tasks are commonly used for graphs with 200 nodes or less (24 / 26 studies, 92\%) and less than 10\% density (25 / 26 studies, 96\%).
	\item Studies with larger graphs ($> 500$ nodes) tend to use multiple types of interaction, while using a single type of interaction is more common in studies with smaller graphs ($\leq 500$ nodes).
	\item The majority of studies with no interaction used graphs with 100 nodes or less (70 / 80 studies, 88\%) and less than 10\% density (57 / 80 studies, 71\%).
	\item The tasks of \emph{Abstract/Elaborate}, \emph{Connect}, \emph{Explore}, \emph{Reconfigure}, and \emph{Select} were used on larger graphs, while \emph{Create}, \emph{Encode}, and \emph{Filter} were used on smaller graphs ($\leq 100$ nodes and $< 10\%$ density).
\end {itemize}

\subsection{Discussion}

\begin{table}
\caption {\label{tab:nodes-size}Four categories of graph size based on number of nodes.
%We only considered values that were reported in the papers.
}

\centering
{\scriptsize
	\begin{tabu} to \columnwidth {cccXX}%
	\toprule 
	& \textbf{\# of Nodes} & \textbf{\# of Studies} & \textbf{\# of Studies with no interaction} & \textbf{\# of Studies with Overview Tasks} \\
	\midrule
	\textbf {small} & $\leq 20$ & 62 (41\%) & 39 (63\%) & 19 (31\%) \\ %49
	\textbf {medium} & $[21, 50]$ & 50 (33\%) & 31 (62\%) & 12 (24\%) \\ %52
	%\textbf {medium} & $[51, 100]$ & 39 & y & Z \\ 
	%\textbf {medium} & $[101, 200]$ & 23 & y & Z \\ 
	\textbf {large} & $[51, 200]$  & 56 (37\%) & 27 (48\%) & 16 (29\%) \\ %63
	\textbf {v. large} & $>200$ & 33 (22\%) & 10 (30\%) & 19 (58\%) \\ %40
\bottomrule
\end{tabu}}
\end{table}

\begin{table}
\caption {\label{tab:density-size}Four categories of graphs based on linear density.  
%We only considered values that were reported in the papers.
}
\centering
{\scriptsize
	\begin{tabu} to \columnwidth {p{1.4cm}ccXX}%
	\toprule 
	& \textbf{Linear Density} & \textbf{\# of Studies} & \textbf{\# of Studies with no interaction} & \textbf{\# of Studies with Overview Tasks} \\
	\midrule
	%\textbf {v. sparse} & $[0, 0.9]$ & 10 & 4 (40\%) & 5 (50\%) \\
	\textbf {tree-like \& disconnected} & $[0, 1.0]$ & 49 (32\%) & 20 (41\%) & 15 (31\%) \\ 
	%\textbf {tree} & $[0.91, 1.0]$ & 37 & 14 (38\%) & 10 (27\%) \\ 
	\textbf {sparse} & $[1.01, 2.0]$ & 67 (44\%) & 40 (60\%) & 20 (30\%) \\ 
	\textbf {dense} & $[2.01,4.0]$  & 31 (20\%) & 18 (58\%) & 7 (23\%) \\ 
	%\textbf {dense} & $>2.01$ & 51 & 25 (49\%) & 18 (35\%) \\ 
	%\textbf {sparse} & $[1.01, 3.0]$ & 87 & 54 (62\%) & 26 (30\%) \\ 
	%\textbf {sparse} & $[1.01, 1.2]$ & 22 & 15 (68\%) & 6 (27\%) \\ 
	%\textbf {sparse} & $[1.01, 1.5]$ & 51 & 31 (61\%) & 16 (31\%) \\ 
	%\textbf {dense} & $[1.21, 1.5]$ & 35 & 22 (63\%) & 13 (37\%) \\ 
	%\textbf {dense} & $[1.51,2.0]$  & 27 & 12 (44\%) & 7 (26\%) \\ 
	%\textbf {v. dense} & $[1.51, 4.0]$ & 58 & 30 (52\%) & 14 (24\%) \\ 
	\textbf {v. dense} & $>4.0$ & 26 (17\%) & 12 (46\%) & 13 (50\%) \\
	%\textbf {dense} & $>3.01$ & 34 & 17 (50\%) & 15 (44\%) \\ 
\bottomrule
\end{tabu}}
\end{table}

While there has been a significant focus on computational scalability of node-link diagrams layout and rendering, in a race to visualise the largest networks, it seems researchers have understudied human cognitive limitations in understanding such diagrams. 
A better knowledge of cognitive scalability in this regard would have several benefits. 
In tools that allow the user to interactively explore a large network through neighbourhood or aggregated views, tool developers could more intelligently control the number of elements in these views. 
Furthermore, if our community could give clear and informed guidance to users of graph visualisation it would help them to select the right tool for their purpose.  For example, in creating figures for papers, biologists reporting on the interactions of particular proteins may be better off showing a focused neighbourhood around those specific proteins rather than providing a hairball.
Similarly, if the users were experimenters, they would choose their corpora, design their tasks and pick layout methods based on these well-defined limits.

Thus, the aim of this survey was to explore factors that would affect cognitive scalability via reviewing existing empirical studies that have used node-link diagrams.
We have noticed that controlled experiments tend to focus on graph datasets of size within a fairly limited window (tens to a few hundred nodes and low density). We also discovered that even though the most common type of tasks performed on a network, in general, is related to the topology of the network, overview tasks become more popular for larger networks. Similarly tasks related to detailed attributes associated to the nodes or edges and browsing are not common on networks with more than a couple hundred nodes. With regards to interaction, studies with large graphs tend to allow for more than one type of interaction. Furthermore, we discovered that some interaction types, such as \emph{Create}, \emph{Encode} and \emph{Filter} are only used on small graphs, while others, e.g., \emph{Abstract/Elaborate}, \emph{Connect}, \emph{Explore}, \emph{Reconfigure}, and \emph{Select} are also used on large graphs.

This survey has also helped identify some weaknesses in the design and reporting of empirical studies that use node-link diagrams. For example, several studies do not report on the sizes of the graphs used, while others do not report on the number of elements visible to the participants at a given time. 
We provide a list of recommendations to overcome these weaknesses in future studies.

%It can also pave way to redefining relative terms, such as small, medium or large graphs into more meaningful representations of 
A motivation for this work was to identify or validate terms that are used to categorise ranges for graph size. Table~\ref{tab:nodes-size} presents four categories with respect to number of nodes. According to the surveyed studies, there are clear cuts at 20, 50, and 200 nodes. We categorise these into ranges that represent small, medium, large and very large graphs. Follwoing significant thresholds of linear density, which have been identified by Melan\c{c}on \cite{melancon2006just}, we categorise sparse, dense, and very dense graphs in Table~\ref{tab:density-size}. Hopefully this breakdown gives future researchers a clear motivation for selecting different graph sizes for their studies. For example, there are only seven studies that use dense graphs with overview tasks. 

Our findings, indicate a threshold at 200 nodes and 10\% density. This threshold is respected by empirical studies that include tasks requiring a detailed analysis of the network. Nonetheless, we believe that this threshold is a result of the expert intuition of the researchers, rather than empirical research. A controlled study is needed to validate and refine this threshold.